\pdfoutput=1
\documentclass[a4paper,prb,showkeys,amsmath,amsfonts,amssymb,superscriptaddress,twocolumn]{revtex4-1}

\usepackage[english]{babel}
\usepackage[utf8]{inputenc}
\usepackage[T1]{fontenc}

\usepackage{graphicx}
\usepackage[caption=false]{subfig}
\usepackage[svgnames]{xcolor}
\usepackage{siunitx}
\usepackage{upgreek}
\usepackage{booktabs}
\usepackage{multirow}
\usepackage{array}
\usepackage[hidelinks]{hyperref}

\bibpunct{[}{]}{,}{n}{}{} 

\begin{document}

\title{Multi-analytical characterization of Fe-rich magnetic inclusions in diamonds}
\author{Marco Piazzi}
\email[Corresponding author. E-mail address: ]{marco.piazzi@unipv.it}
\author{Marta Morana}
\affiliation{Dipartimento di Scienze della Terra e dell'Ambiente, Università di Pavia, Via A. Ferrata 1, I-27100 Pavia, Italy}
\author{Marco Coïsson}
\affiliation{Divisione di Metrologia dei Materiali Innovativi e Scienze della Vita, Istituto Nazionale di Ricerca Metrologica, Strada delle Cacce 91, I-10135 Torino, Italy}
\author{Federica Marone}
\affiliation{Swiss Light Source, Paul Scherrer Institut, 5232 Villigen, Switzerland}
\author{Marcello Campione}
\affiliation{Dipartimento di Scienze dell'Ambiente e della Terra, Università degli Studi di Milano - Bicocca, Piazza della Scienza 4, I-20126 Milano, Italy}
\author{Luca Bindi}
\affiliation{Dipartimento di Scienze della Terra, Università degli Studi di Firenze, Via G. La Pira 4, I-50121 Firenze, Italy}
\author{Adrian P. Jones}
\affiliation{Department of Earth Sciences, University College London, Gower Street, London WC1E 6BT, UK}
\author{Enzo Ferrara}
\affiliation{Divisione di Metrologia dei Materiali Innovativi e Scienze della Vita, Istituto Nazionale di Ricerca Metrologica, Strada delle Cacce 91, I-10135 Torino, Italy}
\author{Matteo Alvaro}
\affiliation{Dipartimento di Scienze della Terra e dell'Ambiente, Università di Pavia, Via A. Ferrata 1, I-27100 Pavia, Italy}

\begin{abstract}
Magnetic mineral inclusions, as iron oxides or sulfides, occur quite rarely in natural diamonds. Nonetheless, they represent a key tool not only to unveil the conditions of formation of host diamonds, but also to get hints about the paleointensity of the geomagnetic field present at times of the Earth's history otherwise not accessible. This possibility is related to their capability to carry a remanent magnetization dependent on their magnetic history. However, comprehensive experimental studies on magnetic inclusions in diamonds have been rarely reported so far. Here we exploit X-ray diffraction, Synchrotron-based X-ray Tomographic Microscopy and Alternating Field Magnetometry to determine the crystallographic, morphological and magnetic properties of ferrimagnetic Fe-oxides entrapped in diamonds coming from Akwatia (Ghana). We exploit the methodology to estimate the natural remanence of the inclusions, associated to the Earth's magnetic field they experienced, and to get insights on the relative time of formation between host and inclusion systems. Furthermore, from the hysteresis loops and First Order Reversal Curves we determine qualitatively the anisotropy, size and domain state configuration of the magnetic grains constituting the inclusions.
\end{abstract}

\keywords{Magnetite; Diamond; X-ray diffraction; Tomography; Alternating Field Magnetometry}

\maketitle

\section{\label{sec:intro} Introduction}
Natural diamonds can provide unique information on the composition and formation processes of the Earth's interior as well as about many fundamental phenomena involved in the geological history of our planet, as for example fluids diffusion into the continental lithosphere \cite{Tappert-diambook,Haggerty-diamond,Bulanova-diamonds, Stachel-diamond,Stachel-diamondbis}. This distinctive feature is related to diamond capability of traveling long distances inside the Earth, from the depth of strata where they were formed, moving towards the surface without being subjected to cracks or breakages. However, it is quite difficult to recover directly from diamonds valuable information about their ages and the pristine thermodynamic conditions and chemico-physical environment present during their growth, because they act as chemically inert materials. In most cases, these conditions and environments are traced back by characterizing and analyzing the properties of mineral inclusions they entrapped \cite{Harris-diaminclus,Shirey-review,Stachel-diaminclusions}, which can reach us almost unaltered thanks to the shielding action of diamonds. For the purpose of restoring this information, many experimental techniques \cite{Nestola-SXRD,Nestola-Mossbauer,Liu-Raman,Nasdala-Raman,Anzolini-Raman} and numerical tools have been developed in the last years allowing to identify the mineral phases composing the inclusions and to relate, by means of analytical equations of state, their crystalline properties to the depth, pressure and temperature of their ancient nucleation and growth processes \cite{Angel-elasticdiammodel,Angel-elasticdiammodelbis,Angel-elasticdiammodeltris,Milani-garnetdiammodel}. A key aspect that should be carefully addressed to avoid incorrect conclusions on this topic is the ascertainment of the time of formation of the inclusion with respect to the time of formation of the host diamond. Syngenetic inclusions, contrary to proto- and epigenetic ones, are indeed the only ones providing us with accurate information about the environment of growth of the diamonds since they nucleate simultaneously with their host. Some criteria, developed quite recently and looking at the relative lattice orientation between the host and the inclusion structures \cite{Nestola-olivinediam, Nestola-syninclusions,Nimis-orientations} or at the presence of fractures into diamonds \cite{Meyer-fractures,Bulanova-fractures}, rather than at the morphology imposed by the host only \cite{Harris-syninclus}, resulted to be effective in distinguishing between the proto-/syngenetic and the epigenetic class of inclusions.

In this context, very few studies have been reported so far about the magnetic properties of mineral inclusions found either in kimberlitic diamonds \cite{Clement-magnincl}, in carbonados \cite{Kletetschka-magncarbonado,Coey-magncarbonado} or in mixtures of kimberlite-source polycrystalline diamonds and carbonados \cite{Collinson-magnpolydiam}. These inclusions have been identified as iron sulfides (pyrrhotite) entrapped in the deeper parts of the diamonds in Ref.~\cite{Clement-magnincl} and as iron oxides (magnetite) for the kimberlite-source polycrystalline diamonds studied in Ref.~\cite{Collinson-magnpolydiam}, while Refs.~\cite{Kletetschka-magncarbonado,Coey-magncarbonado} have evidenced that the magnetic carriers responsible for the detected ferromagnetic signals are present at the surface or in open pores, rather than in the bulk of the examined carbonados. This kind of inclusions is able to carry a natural remanent magnetization (NRM) which represents a signature of the geomagnetic field they have been subjected to along their history. Therefore, magnetic inclusions can be regarded as geological objects of great interest since they might bring useful data about the growth conditions of the host system \cite{Gilder-pressureFMPMtrans} and, at the same time, about the paleointensity of the geomagnetic field present in key geological eras, which would be otherwise not accessible. It is worth noting that the latter information may be hindered if the inclusion has strong magnetic anisotropy, as for example may happen when dealing with pyrrhotite \cite{Clement-magnincl}. In particular, the work by Clement et al. \cite{Clement-magnincl} offers a complete study about the magnetic characterization of pyrrhotite inclusions entrapped in eleven diamonds of millimeter size coming from the Orapa kimberlite mine in Botswana. The authors propose an experimental procedure to ascertain the main magnetic properties of the inclusions based on: (\textit{i}) progressive alternating field demagnetization and isothermal remanent magnetization (IRM) measurements, aimed at establishing the natural remanent magnetization carried by the system; (\textit{ii}) thermal demagnetization measurements, with the purpose of determining the Curie temperature of the inclusions and their possible chemical alteration due to heating processes; (\textit{iii}) hysteresis loops acquisition at different orientations, to investigate the coercive field, remanence and potential magnetic anisotropy present in the system. The presence of quite regular shapes and distinct features in their diamonds allowed the authors to perform a visual identification of the inclusions. When present, such distinct features, as for example cleavage planes and well visible fractures connected to the diamond surface and surrounding the inclusions, allows also to visually distinguish epigenetic inclusions from proto-/syngenetic ones. Unfortunately, in some cases natural diamonds do not show these clear features and other analytical tools (e.g. optical microscopy, X-ray diffraction, ...) are required to clearly identify the inclusions and to determine their properties. As shown in Ref.~\cite{Clement-magnincl}, thermal demagnetization can help in identifying magnetic minerals through their Curie temperatures, but this process may bring to undesired chemical alteration of the pristine inclusions. Furthermore, the identification of a magnetic mineral only through thermal demagnetization can be hard when dealing with multiple magnetic inclusions, characterized by similar Curie temperatures, within a single diamond. 

In this work we thus suggest an experimental procedure allowing to partially overcome the above mentioned difficulties and we apply it to investigate the crystallographic, morphological and magnetic properties, not accessible by optical and visual means alone, of iron oxides inclusions entrapped in a series of single-crystal diamonds coming from Akwatia (Ghana). This procedure exploits several non-destructive, efficient and relatively fast techniques detailed in Sec.~\ref{sec:methods}, i.e. X-ray diffraction (XRD), Synchrotron-based X-ray Tomographic Microscopy (SRXTM) and Alternating Gradient Field Magnetometry. In particular, XRD allows to determine the crystal structure of the inclusions in a repeatable way without altering them \cite{Nestola-XRDtomo,Pearson-XRD,Nestola-XRDolivine,Angel-XRDhighP}, while SRXTM is performed to establish which inclusions can be regarded as epigenetic by searching for the potential presence of fractures connecting them to the surface all along the diamond interiors \cite{Nimis-tomo}. SRXTM allows also to give a reasonable estimate of the linear size, volume and shape of the host-inclusion system, which is an important piece of information to determine more precisely the magnetization of the inclusions or to perform numerical calculations. Finally, Alternating Gradient Field Magnetometry measurements represent a quick and enough sensitive tool to investigate the magnetic properties of the system. These properties may provide further constraints when attempting to unveil the thermodynamic conditions of growth of the inclusions, because of the pressure-temperature dependence of the magnetic response of any material with magnetic order. In Sec.~\ref{sec:results} we present the experimental results obtained, while in Sec.~\ref{sec:discussion} we propose their possible interpretation. Sec.~\ref{sec:conclusions} is finally devoted to draw conclusions, proposing also some possible routes for future works on this topic.                    

\section{\label{sec:methods} Samples and methods}
\begin{figure}[htbp]
\centering
\subfloat[][\label{subfig:CAST2-container}]{\includegraphics[width=0.22\textwidth]{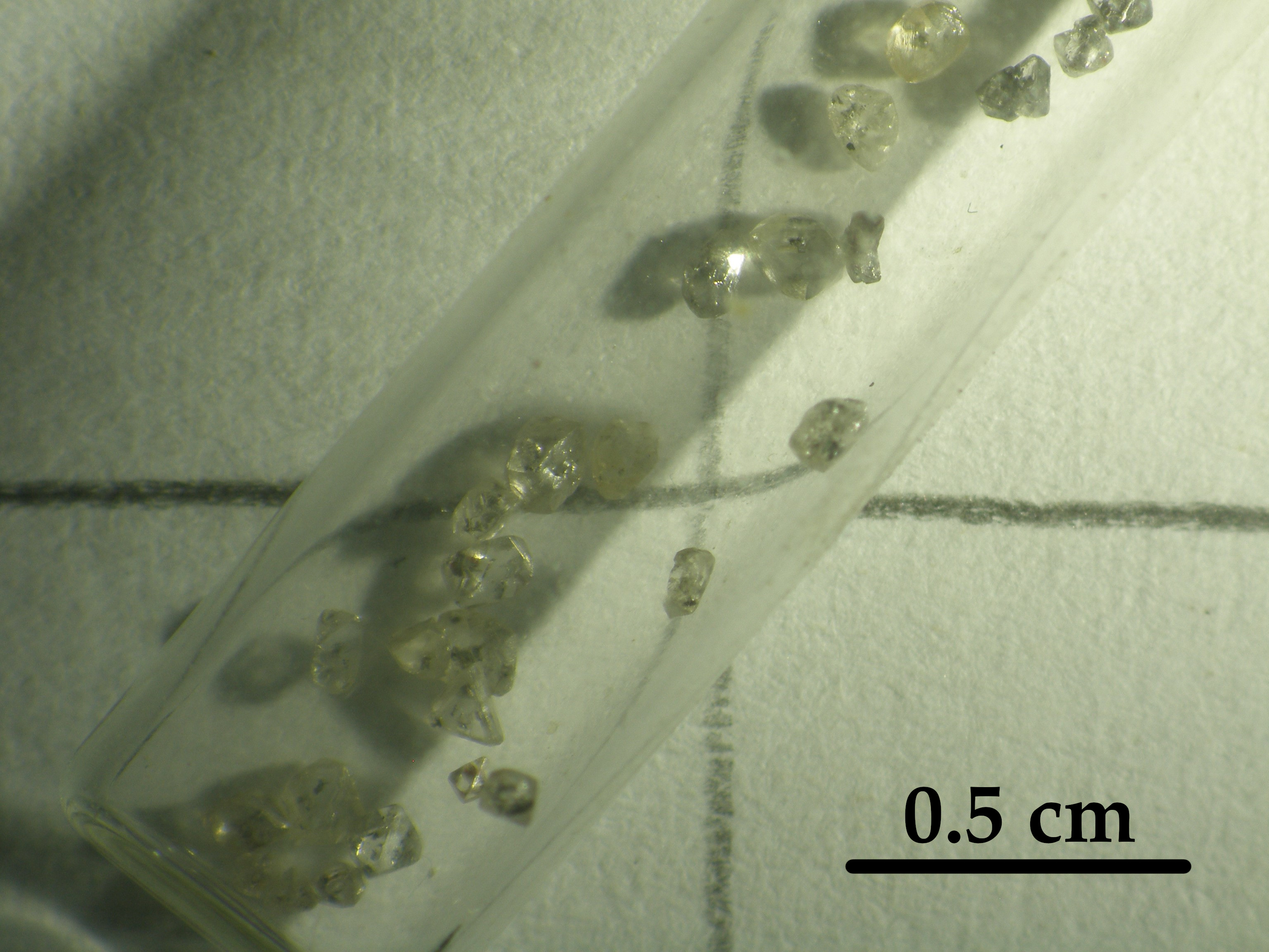}}\; 
\subfloat[][\label{subfig:CAST2-1-opt}]{\includegraphics[width=0.222\textwidth]{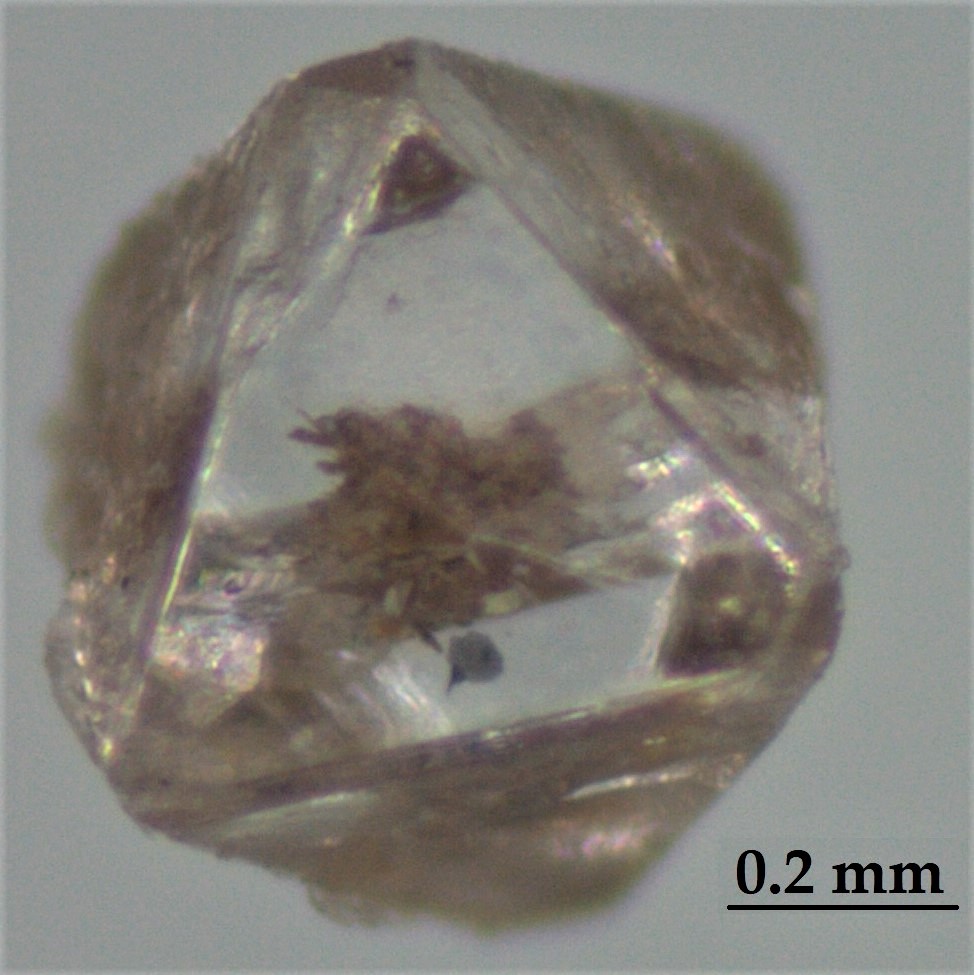}}\\ 
\subfloat[][\label{subfig:CAST2-5-opt}]{\includegraphics[width=0.22\textwidth]{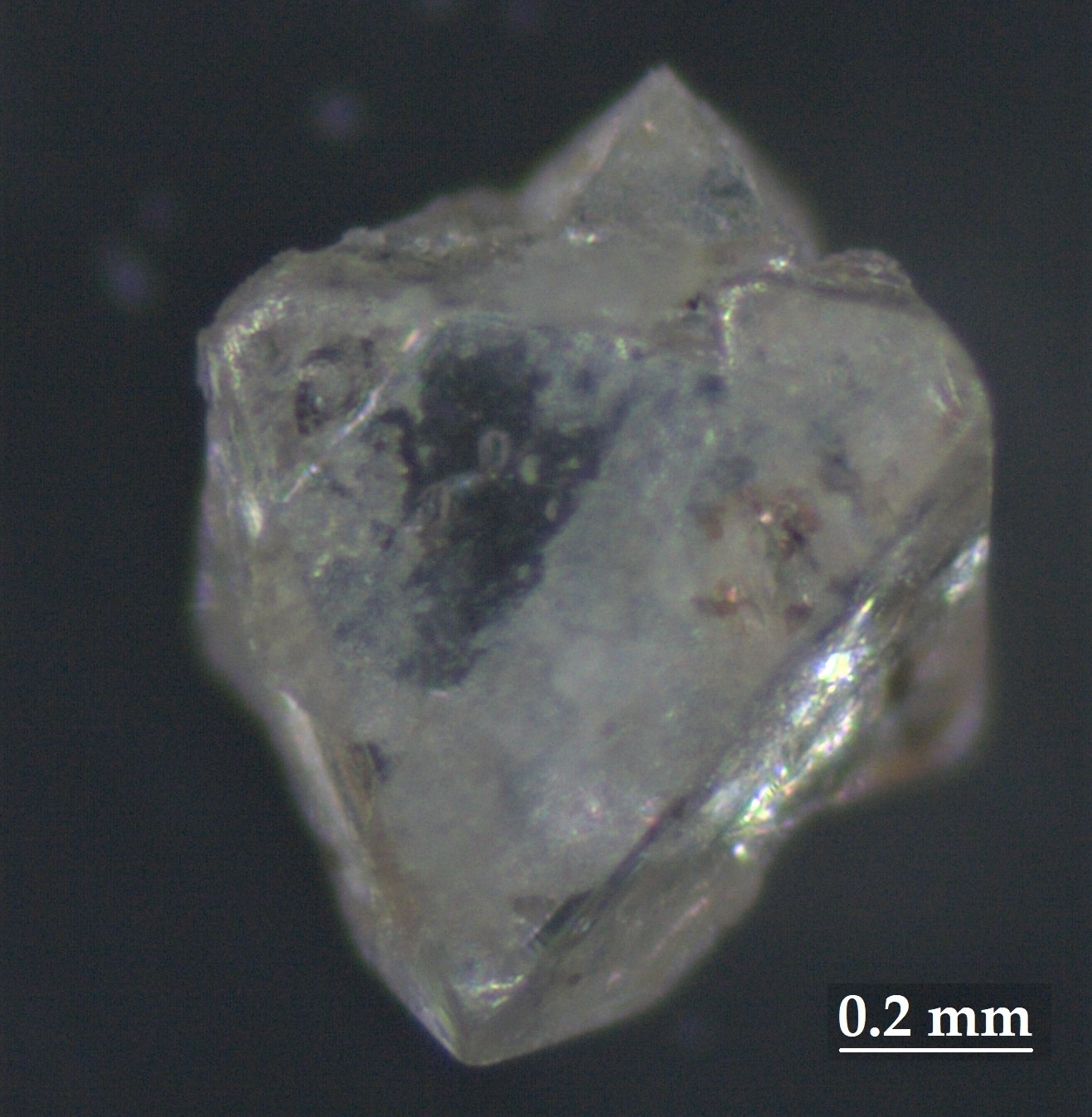}}\; 
\subfloat[][\label{subfig:CAST2-13-opt}]{\includegraphics[width=0.222\textwidth]{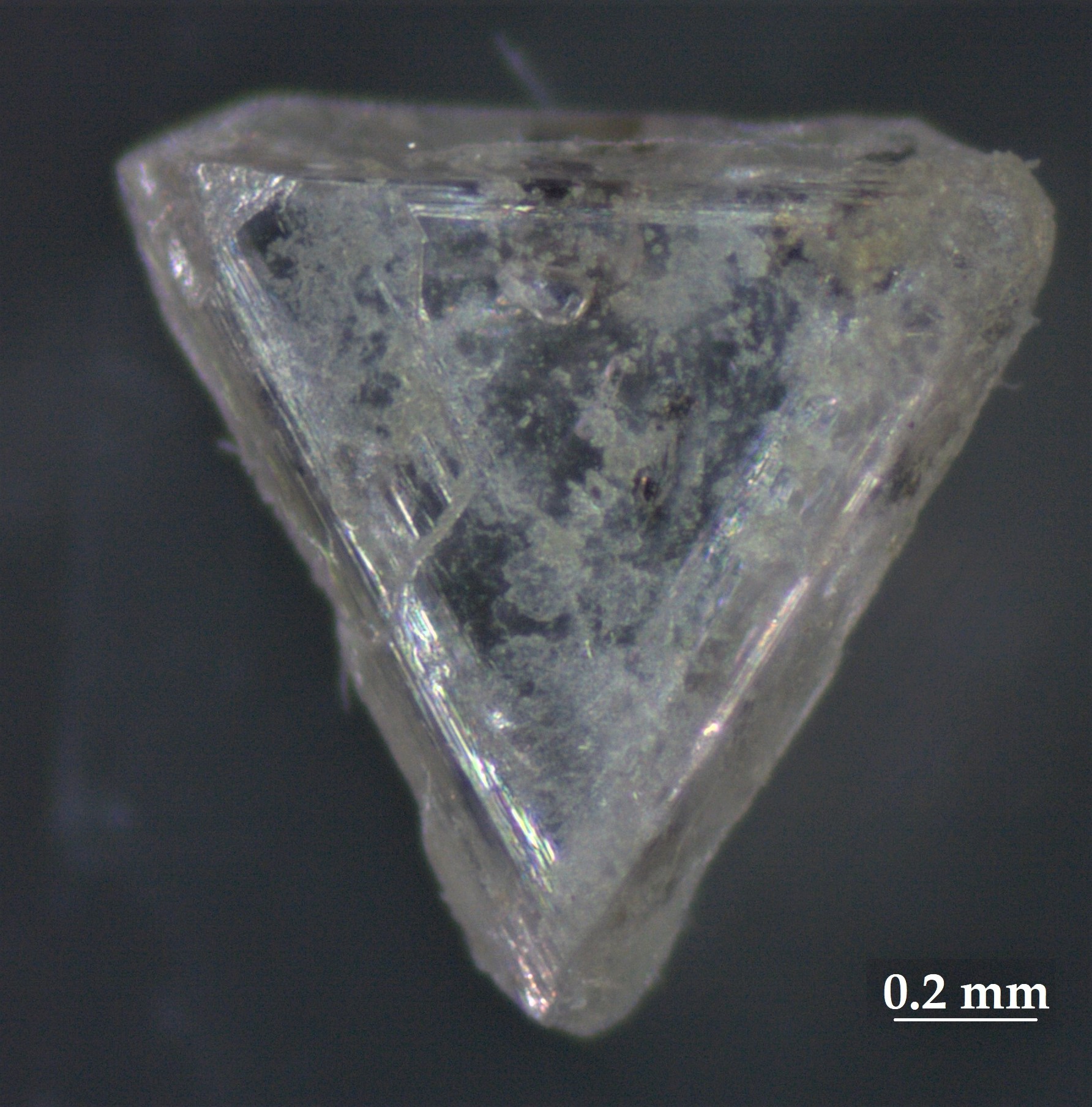}} 
\caption{(\textit{a}) Diamonds belonging to the CAST2 series and optical images of: (\textit{b}) CAST2-1; (\textit{c}) CAST2-5; (\textit{d}) CAST2-13 samples chosen as examples. Diamonds show quite different shapes; inclusions are in most cases distinguishable by their dark-gray, black color.} 
\label{fig:CAST2-diamonds}
\end{figure}

The series of diamonds investigated in our study, labelled as CAST2 (\figurename~\ref{subfig:CAST2-container}) and provided us as a courtesy of Dr. H. J. Milledge from the University College of London, were collected in the Birim River valley of Akwatia, in Ghana, in 1960-1970. Unfortunately, it was not possible to recover the exact orientation of the samples when they were extracted from the deposit, since alluvial diamonds are randomly oriented in the alluvial valley. \num{15} diamonds of the series, clearly showing the presence of inclusions, have been selected through optical microscopy for subsequent measurements. Selected diamonds show quite different shapes (\figurename~\ref{subfig:CAST2-1-opt}, \ref{subfig:CAST2-5-opt}, \ref{subfig:CAST2-13-opt}), from clearly octahedral to irregular habit, and range in size between $\sim$\SI{0.5}{\milli\metre} and \SI{1.5}{\milli\metre}. Inclusions appear in most cases dark-gray or black in color, although a clear identification of their distinctive features is made difficult by the many reflexes present in diamonds and by the unclean surface. 

To identify the mineral phases present as inclusions, estimate their sizes and unveil the possible presence of fractures in the diamond host, all samples have been analyzed combining XRD and SRXTM. \num{4} diamonds (CAST2-1, -6, -7, -12) containing inclusions that we could identify as magnetic have also been chosen to undergo magnetic characterization with an Alternating Gradient Force Magnetometer (AGFM) \cite{Flanders-AGFM}, by acquiring IRM and backfield curves, and complete hysteresis loops and First Order Reversal Curves (FORCs). The masses of the samples have been measured to be \SI{0.48}{\milli\gram}, \SI{2.32}{\milli\gram}, \SI{1.27}{\milli\gram}, \SI{1.85}{\milli\gram}, respectively, with an uncertainty of $\sim 2\%$. The combined knowledge of these magnetic features should be sufficient for determining with high enough accuracy the natural remanent magnetization of the inclusions and to interpret their magnetic behaviour in terms of the magnetic granulometry (single- or multi-domain state) of their particles. Such interpretation, combined with the appreciation of magnetic anisotropy, could permit to evaluate the intensity of the geomagnetic field to which inclusions have been subjected to. This comparison actually goes beyond the scope of the present paper and will be the subject of future works. 

\subsection{\label{subsec:XRD} XRD acquisitions} 
XRD measurements have been performed with a Rigaku-Oxford Supernova single-crystal diffractometer. The instrument mounts a Dectris Pilatus3 R 200K-A detector and it is equipped with a molybdenum microfocus source ($\lambda_\text{Mo}\simeq$ \SI{0.71}{\angstrom}) and 4-circles K geometry. Diamonds were attached on brass pins with wax and mounted on a goniometer head allowing for centering the crystal onto the incoming X-ray beam (\figurename~\ref{subfig:sample-XRD}) and $\varphi$ scans of few degrees, usually $\sim\,$\SIrange[range-phrase=--]{40}{50}{\degree} in steps of \SI{0.5}{\degree}, have been performed around each selected inclusion. The most significant frames acquired in each scan have been first corrected by masking the diffraction spots corresponding to the host diamond and then integrated with CrysAlis$^\text{Pro}$ software \cite{Crysalis}, and finally merged together by summing them up with the help of HighScore software \cite{Highscore}.    

\subsection{\label{subsec:microtomo} SRXTM scans}
Synchrotron radiation absorption-based tomographic microscopy has been performed at the TOMCAT-X02DA beamline at the Swiss Light Source facility of the Paul Scherrer Institut \cite{Stampanoni-TOMCAT}. 2D radiographic projections of the samples have been collected by setting the X-ray beam energy to \SI{20}{\kilo\electronvolt} and by using: (\textit{i}) a \SI{5.8}{\micro\metre} thick LSO:Tb scintillator; (\textit{ii}) an Optique Peter high-resolution microscope accommodating 10x, 20x and 40x Olympus UPLAPO objectives; and (\textit{iii}) a high sensitive, low noise, large field-of-view pco.Edge 5.5 optical camera, featuring a sensor size of \num{2560} $\times$ \num{2160} pixels with a pitch size of \SI{6.5}{\um}. Tomographic volumes have been reconstructed at the facility by means of a highly optimized software based on Fourier transform algorithms \cite{Marone-tomoalgo}. Obtained 3D reconstructed volumes consist of \num{2160} slices each, with a spatial resolution of $\sim\,$\SIrange[range-phrase=--]{0.5}{2}{\um} depending on the objective used. Reconstructed 3D images have been subsequently post-processed with Thermo Scientific\texttrademark\ Avizo\texttrademark\ software to get an estimate of the shape, size and volume of the diamonds and their inclusions.           

\subsection{\label{subsec:AGFM} AGFM meaurements}
\begin{figure}[htbp]
\centering
\subfloat[][\label{subfig:sample-XRD}]{\includegraphics[width=0.26\textwidth]{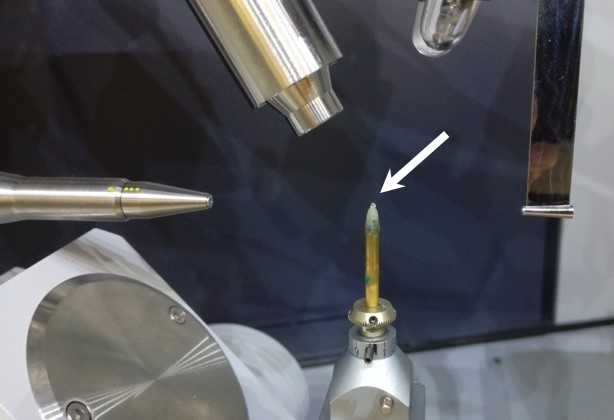}}\;
\subfloat[][\label{subfig:sample-AGFM}]{\includegraphics[width=0.2\textwidth]{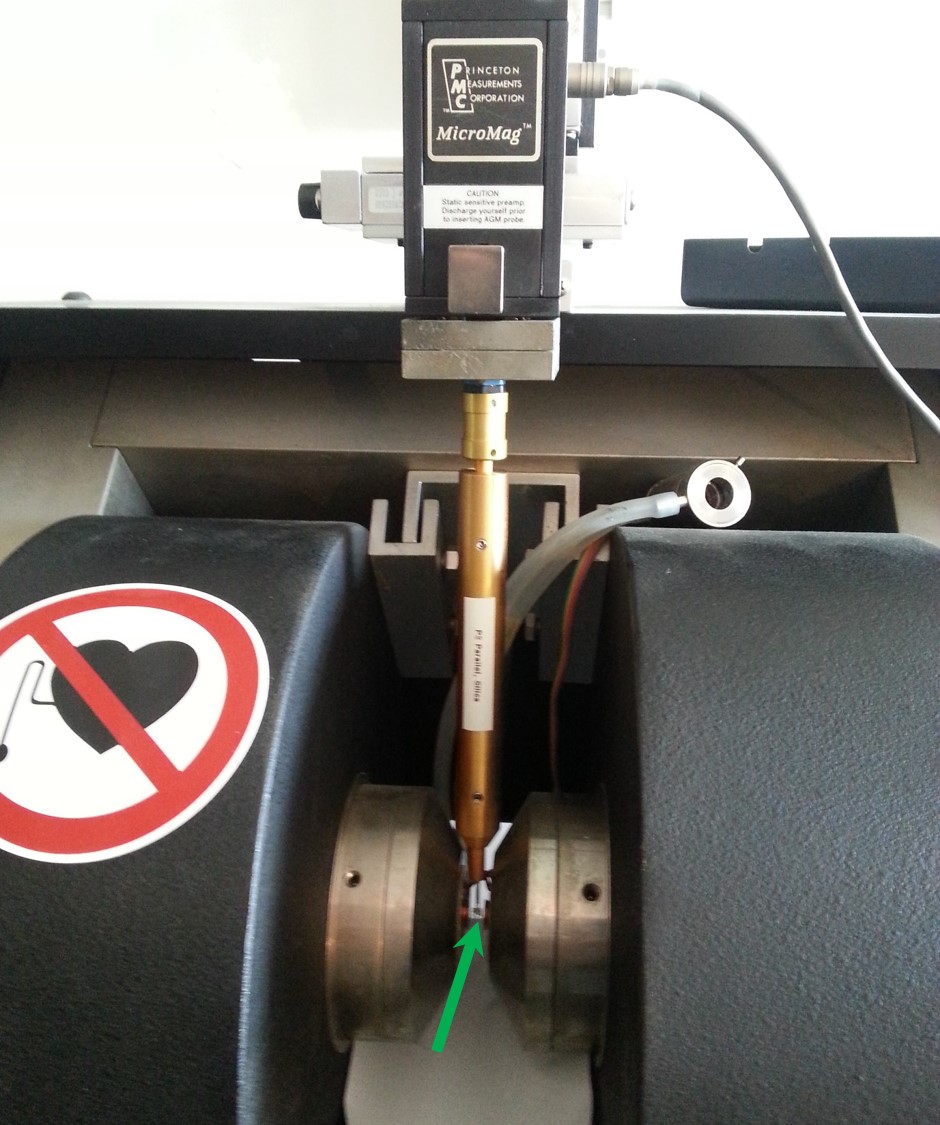}}
\caption{Setup of (\textit{a}) diffractometer and (\textit{b}) AGFM used for crystallographic and magnetic characterization. Diamonds are mounted on a pin in the diffractometer and on a rod in the AGFM, indicated with arrows in the pictures.}
\label{fig:sample-instrum}
\end{figure}

Magnetic characterization of inclusions, as detected in XRD measurements, has been performed with a Lake Shore Cryotronics MicroMag 2900 AGFM. The instrument allows to measure the scalar component of the magnetic moment of a specimen, along the direction of an applied, uniform field $H$, with a nominal sensitivity of \SI{e-11}{\ampere\metre\squared} and a resolution that in our acquisitions varied between $\sim\,$\SI{e-11}{\ampere\metre\squared} and $\sim\,$\SI{2.5e-11}{\ampere\metre\squared} depending on the sample. The uniform and constant flux intensity of the applied field $\mu_0 H$, with $\mu_0=4\uppi\cdot\num{e-7}$ \si[per-mode=symbol]{\weber\per\ampere\per\metre} being the permeability of vacuum, can vary in the range \SIrange[range-phrase={, },range-units=brackets]{-2.2}{2.2}{\tesla} in minimum steps of $\sim\,$\SI{0.05e-3}{\tesla}, although the flux intensity of the field needed in our experiments to achieve magnetic saturation of the samples did not exceed \SI{1}{\tesla}. Throughout the measurements, specimens have been centered into the gap of the 2-probes electromagnet generating $H$ by means of a rod mounted on a piezoelectric sensor (\figurename~\ref{subfig:sample-AGFM}). Acquisitions have been performed by applying a weak, non-uniform, alternating field $H_\text{ac}$ in the $x$-direction longitudinal to $H$ and setting $\partial H_\text{ac}/\partial x=$ \SI[per-mode=symbol]{1.5}{\tesla\per\meter}.

\begin{figure}[htbp]
\centering
\includegraphics[width=0.46\textwidth]{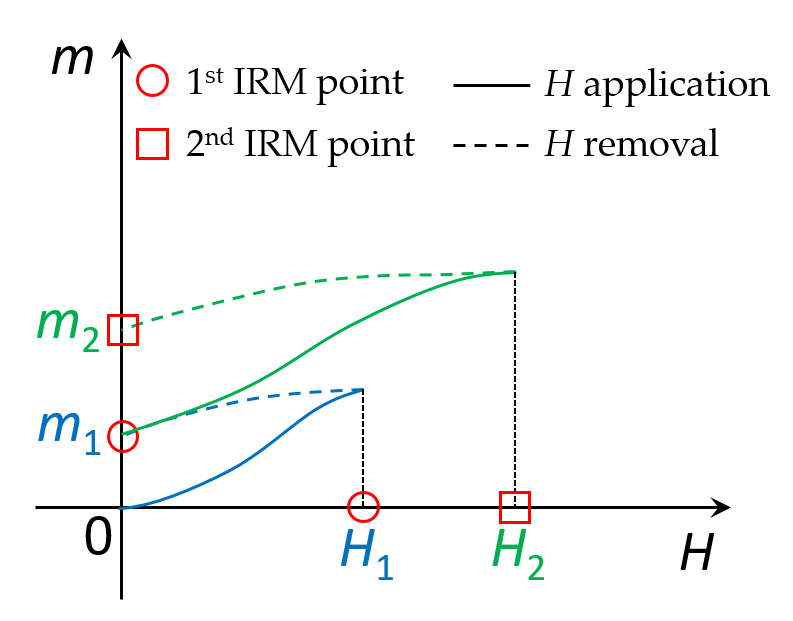}
\caption{Sketch of the procedure followed to acquire IRM curves. Magnetic field $H$ (the uniform, stronger field in the AGFM) is successively applied at increased values $H_1 < H_2 < \ldots$ (solid lines) and subsequently removed (dashed lines). The corresponding components of the remanent magnetic moment (or of the magnetization) $m_1, m_2, \ldots$, along the direction of $H$, are collected for each field value only after its removal.}
\label{fig:IRM-sketch}
\end{figure}

For each selected diamond we acquired IRM and backfield curves, major hysteresis loops and full sets of FORCs. As schematically depicted in \figurename~\ref{fig:IRM-sketch}, IRM curves are obtained by measuring the remanent magnetic moment of the sample $m_\text{IRM}(H)$ after application and subsequent removal of increasing fields $H$, starting from $H=0$. It is worth noting that since $m_{\text{IRM}}(0)$, i.e. the first point of the IRM curve, depends on the entire magnetic history experienced by the sample prior to any magnetic treatment and it is irreversibly cancelled after measurement, its acquisition is a crucial step and it has to be accomplished prior to any other measurements or treatments of the samples. Backfield curves are acquired similarly to IRM ones but in this case decreasing the applied field from the saturating value downward to $H=0$, applying then small reverse, negative fields of increasing absolute intensity, and finally going back to $H = 0$, when the value of the remanent moment is measured. FORCs represent a full set of minor branches, lying inside the unique major hysteresis loop characterizing a given system, collected with a conventional procedure. Starting from the positive saturation field $H_\text{sat}$, above which the magnetic moment is essentially constant, the magnetic field $H$ is decreased down to a reversal point $-H_\text{sat} \leq H_\text{rev} \leq  H_\text{sat}$ and then increased up again to $H_\text{sat}$. Magnetic moment (or magnetization) values $m(H; H_\text{rev})$ are recorded along the ascending branch of the loop $H_\text{sat} \rightarrow H_\text{rev} \rightarrow H_\text{sat}$, therefore at all magnetic fields $H_\text{rev} \leq H \leq H_\text{sat}$.

\section{\label{sec:results} Results}

\subsection{\label{subsec:XRD-res} XRD}
The XRD analysis shows that the inclusions in the \num{15} selected diamonds comprise different mineral phases, that are present as both single crystal and polycrystalline material. Only in two samples, CAST2-4 and CAST2-8, no magnetic phases have been detected, while in the others both magnetic and non-magnetic phases were present. Among the non-magnetic minerals, single-crystal olivine has been identified in most inclusions, while the presence of single-crystal quartz and garnet has been more sporadically detected. Focusing on magnetic phases, in most inclusions they have been identified as magnetite (Fe$_3$O$_4$) or magnesioferrite (MgFe$_2$O$_4$), which are ferrimagnetic materials, and in few cases as hematite ($\alpha$-Fe$_2$O$_3$), a canted antiferromagnet. In particular, the latter, if present, has been always found in a mixture with magnetite/magnesioferrite. Furthermore, all Fe-rich phases generally appeared as powders and only in one diamond (CAST2-7) single-crystal magnetite has also been detected. Finally, it is important to note that magnetite and magnesioferrite have very similar lattice parameters and spinel structure, reflected in similar XRD patterns that make nearly impossible their separated identification especially if investigated as inclusions in diamonds (see Ref.~\cite{Angel-centurypaper}). For this reason, in what follows we will refer indistinctly to magnetite or magnesioferrite whenever this kind of phase is involved.    

\begin{figure}[htbp]
\centering
\subfloat[][\label{subfig:CAST2-1-XRD}]{\includegraphics[width=0.475\textwidth]{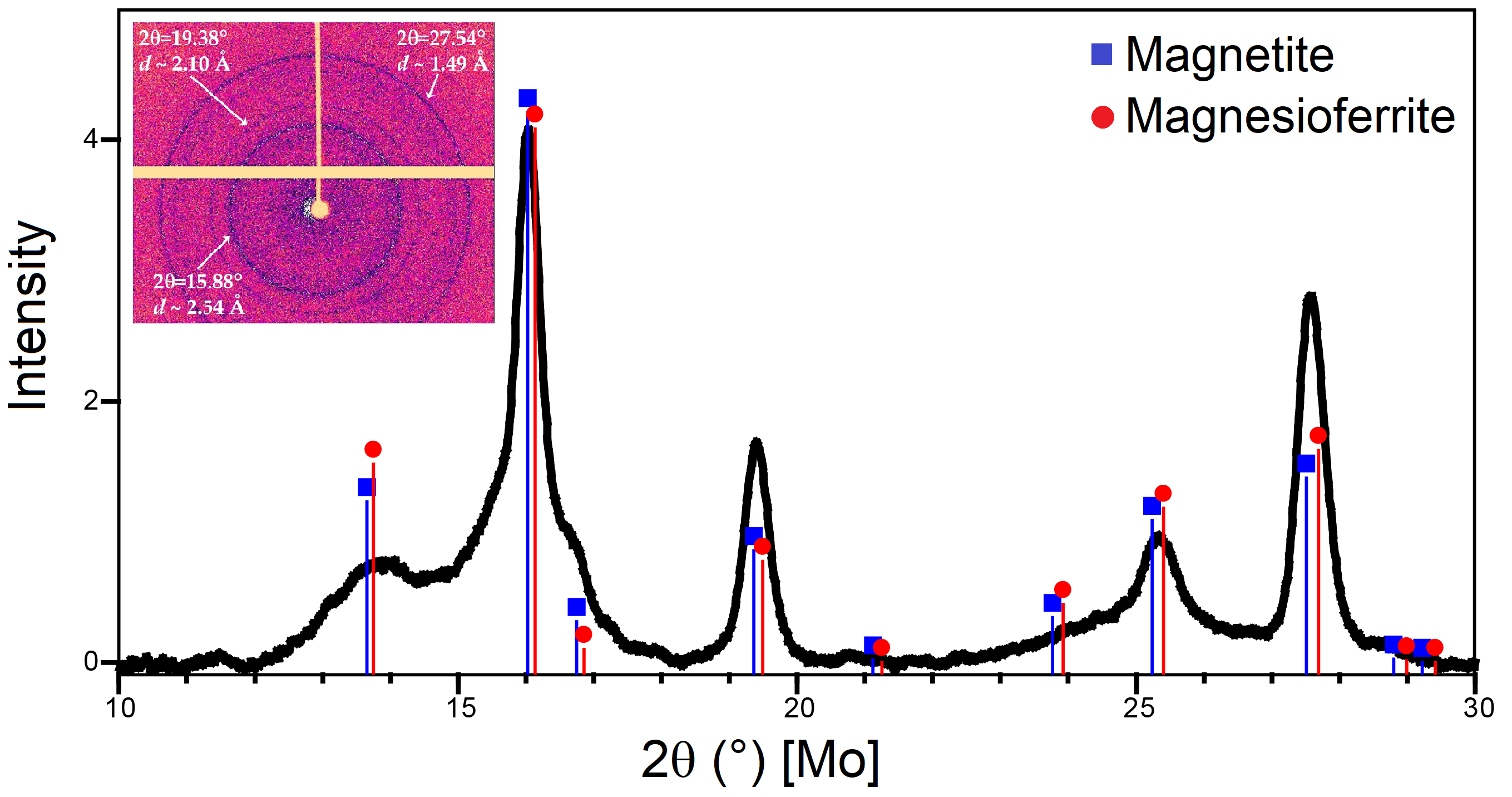}}\\
\subfloat[][\label{subfig:CAST2-6-XRD}]{\includegraphics[width=0.47\textwidth]{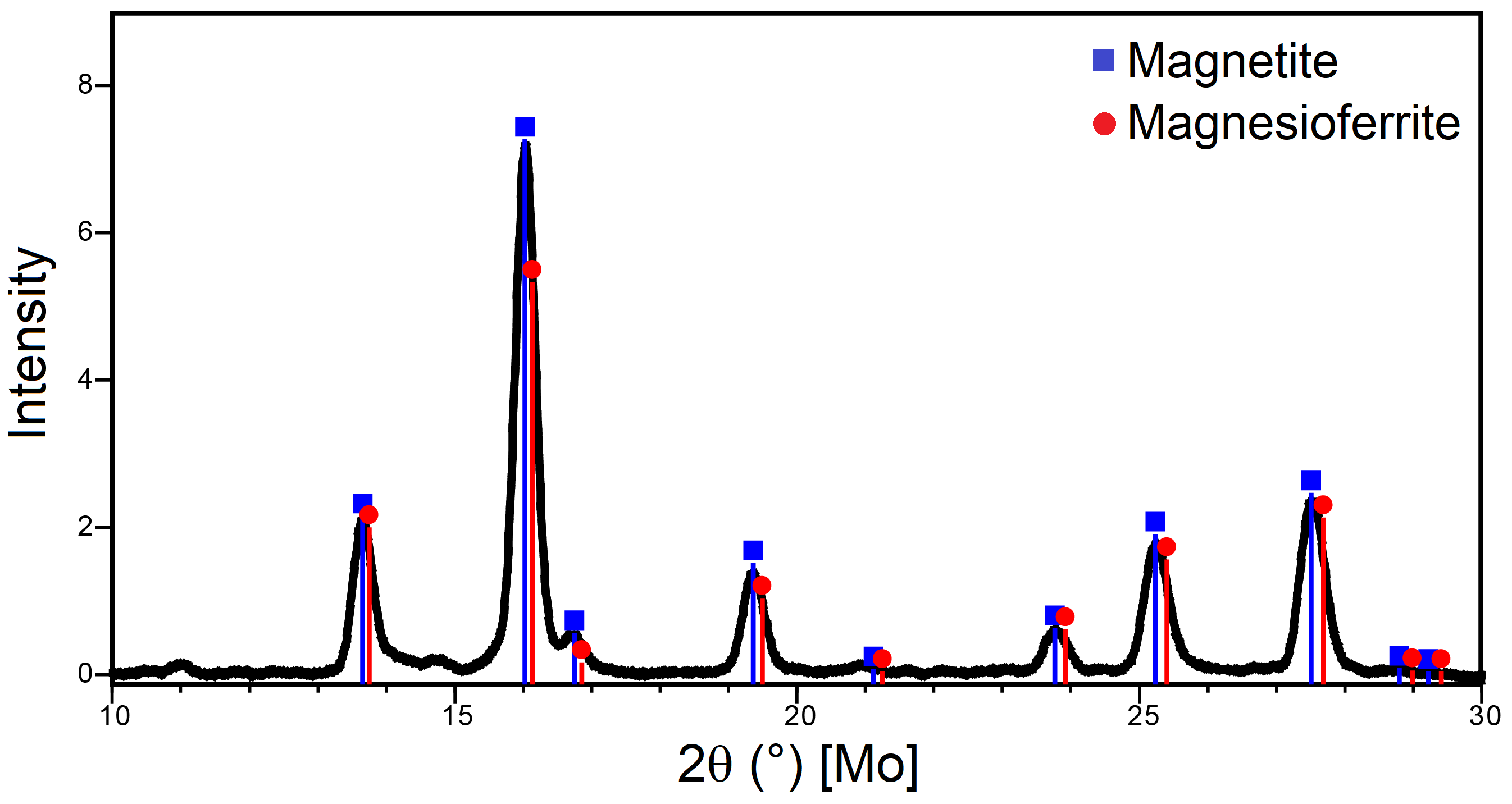}}\\
\subfloat[][\label{subfig:CAST2-7-XRD}]{\includegraphics[width=0.47\textwidth]{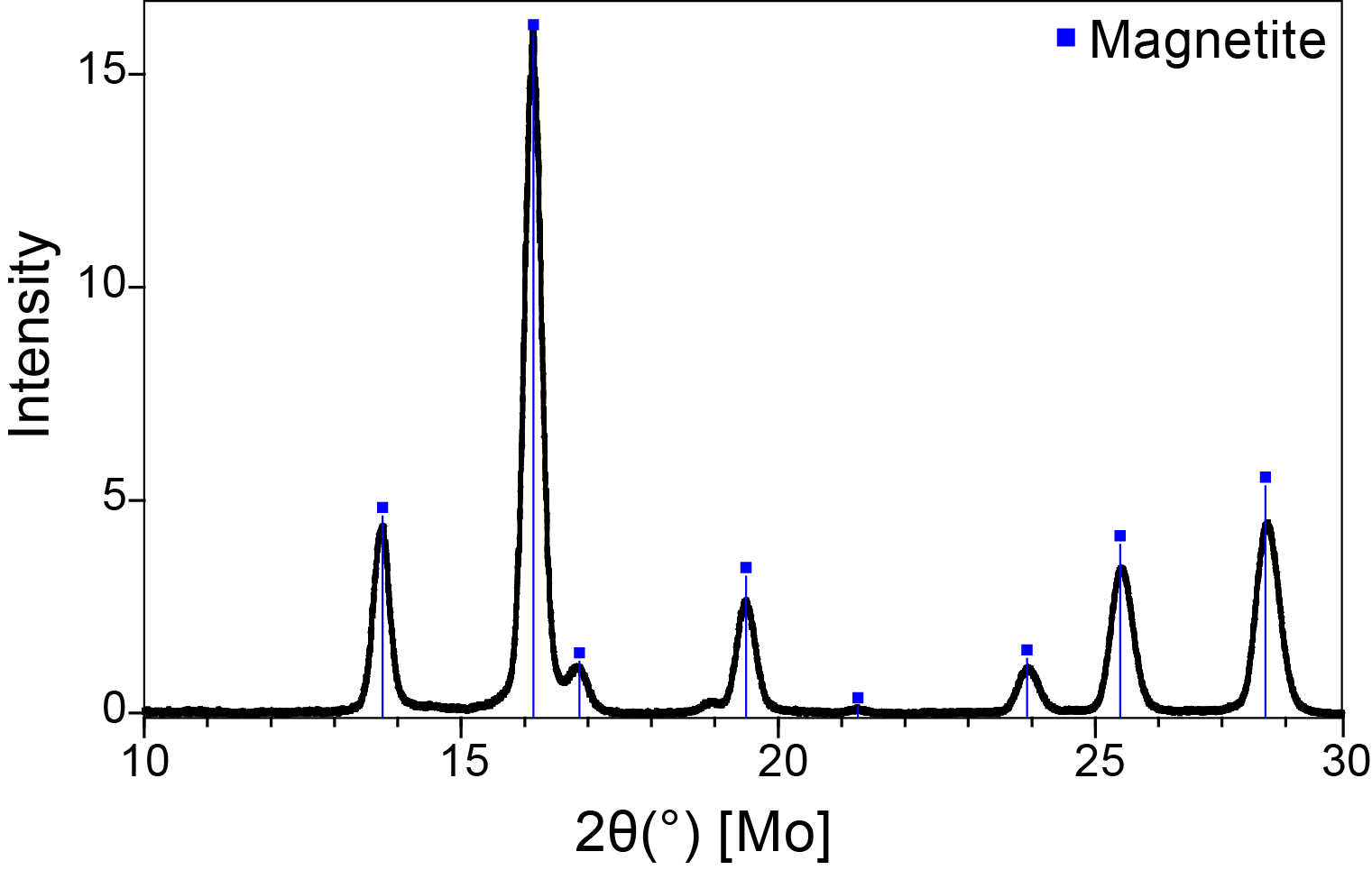}}\\
\subfloat[][\label{subfig:CAST2-12-4-XRD}]{\includegraphics[width=0.47\textwidth]{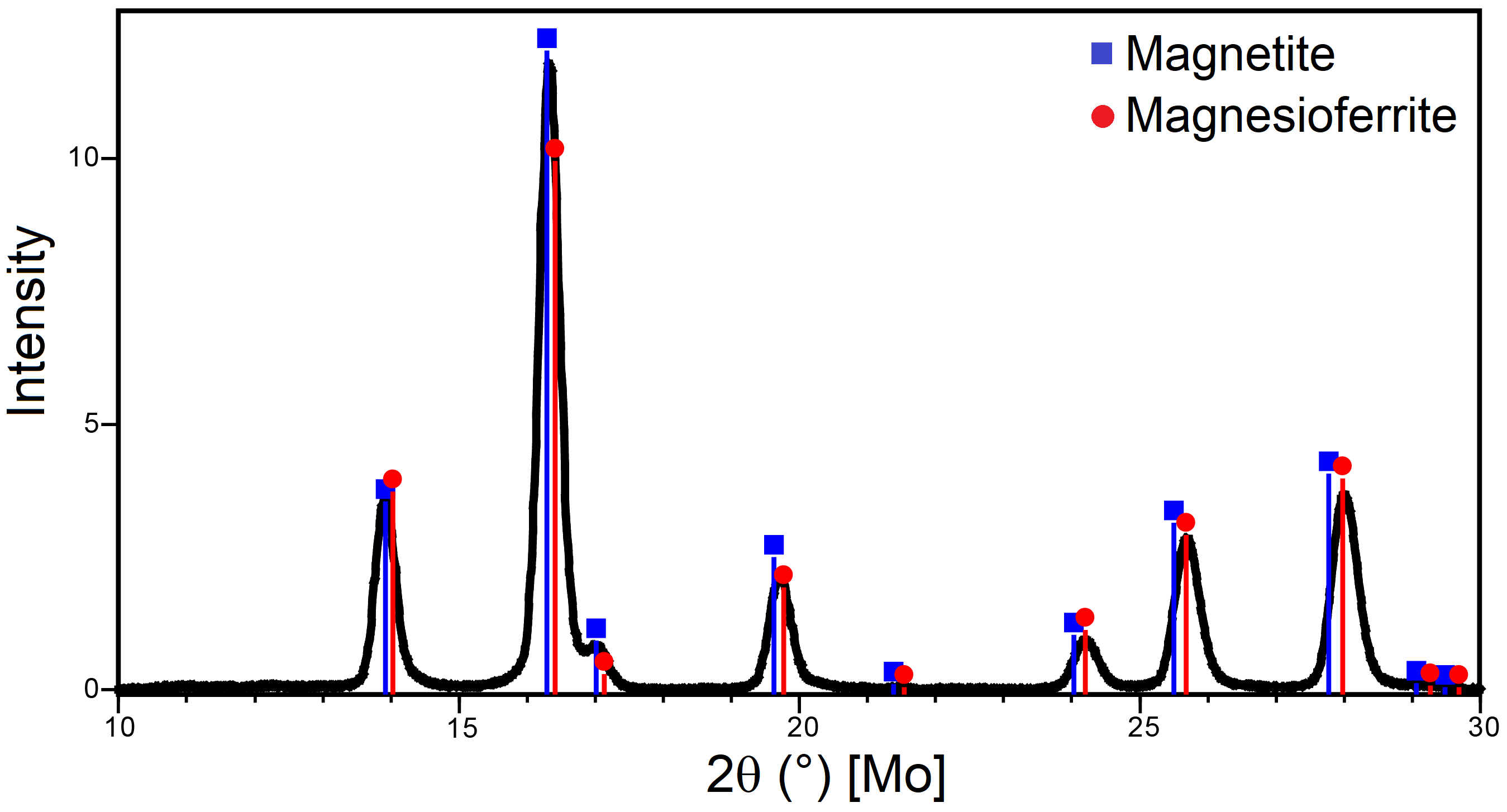}}
\caption{Diffractograms identifying magnetic phases (magnetite -- blue squares; magnesioferrite -- red circles) in: (\textit{a}) CAST2-1; (\textit{b}) CAST2-6; (\textit{c}) CAST2-7 and (\textit{d}) CAST2-12 diamonds. Diffractograms related to non-magnetic inclusions are not reported. Inset in \figurename~\ref{subfig:CAST2-1-XRD} shows the pattern from an acquisition at fixed $\varphi$; diffraction spots related to diamond are not shown. Phase identification performed with the HighScore software.}
\label{fig:CAST2-XRD}
\end{figure}

In particular, for our purposes a promising sample shall contain a unique magnetic phase, not necessarily derived from a single inclusion, in order to surely identify the source of the magnetic signal. The additional presence of non-magnetic minerals does not constitute a problem as they would only contribute with a weak diamagnetic response superimposed to that of the diamond, that is why diffractograms related to non-magnetic inclusions present in the samples are not reported here. According to the above reasoning, CAST2-1 represents the most interesting diamond in the suite, since it contains a unique polycrystalline magnetic phase, as clearly shown from the rings present in the pattern reported in the inset of \figurename~\ref{subfig:CAST2-1-XRD}. Beyond the CAST2-1 diamond, we have selected three more samples, i.e. CAST2-6, CAST2-7 and CAST2-12, to perform magnetic characterization. They contain multiple inclusions, some of them comprising only non-magnetic phases, while others possessing also a magnetic character. Diffractograms corresponding to the magnetic inclusions present in the \num{4} diamonds are reported in \figurename~\ref{fig:CAST2-XRD} and show that all the inclusions have been identified as magnetite or magnesioferrite.

\subsection{\label{subsec:tomo-res} SRXTM}
\begin{figure}[htbp]
\centering
\subfloat[][\label{subfig:tomo-CAST2-1}]{\includegraphics[width=0.46\textwidth]{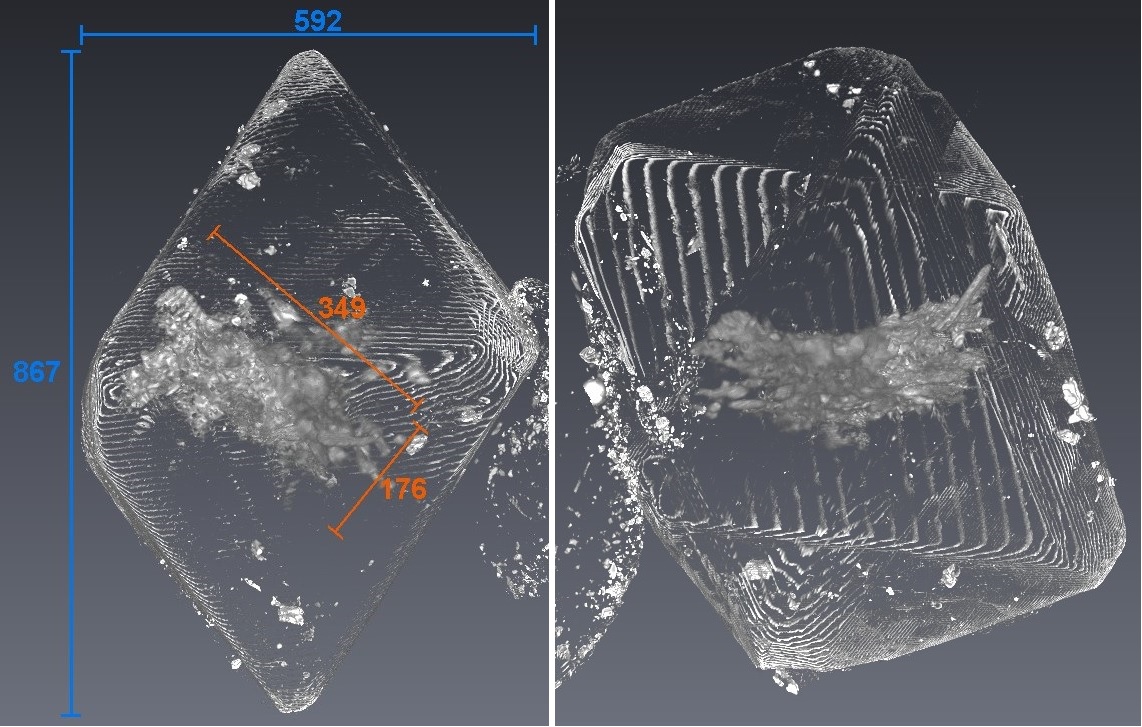}}\\
\subfloat[][\label{subfig:tomo-CAST2-6}]{\includegraphics[width=0.46\textwidth]{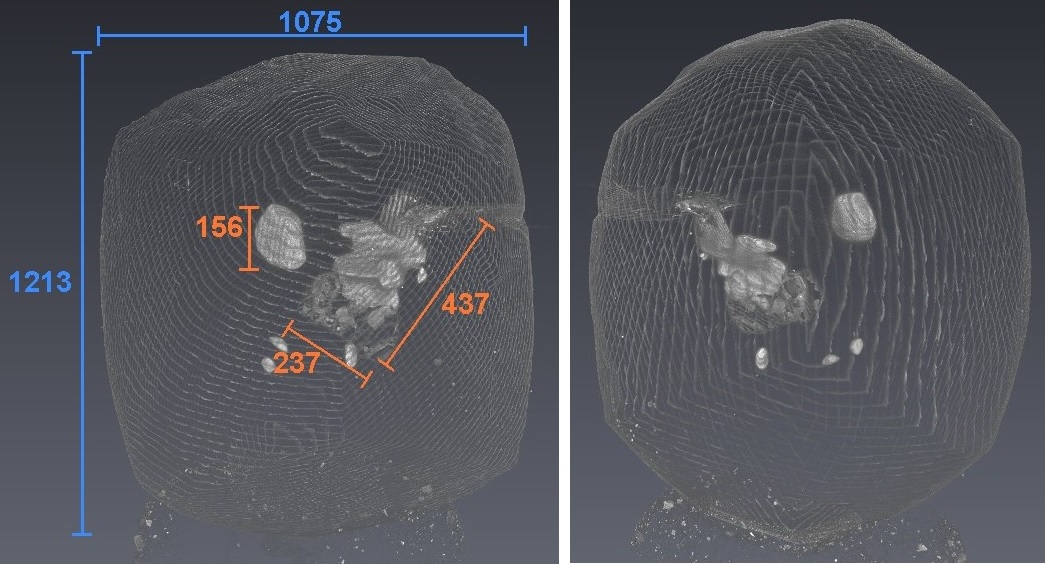}}\\
\subfloat[][\label{subfig:tomo-CAST2-7}]{\includegraphics[width=0.46\textwidth]{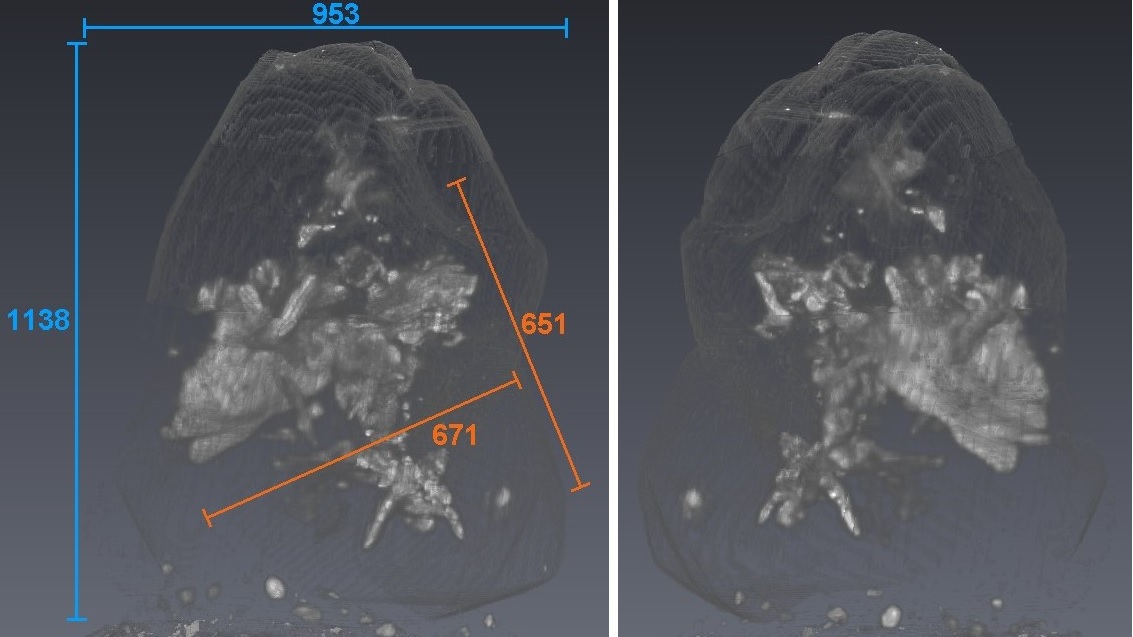}}\\
\subfloat[][\label{subfig:tomo-CAST2-12}]{\includegraphics[width=0.455\textwidth]{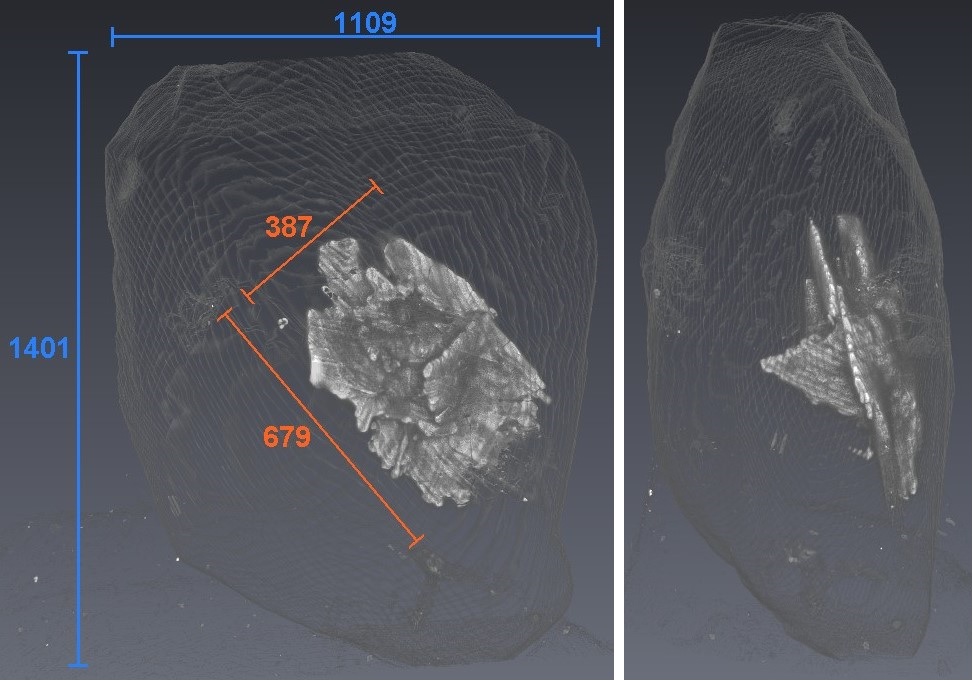}}
\caption{3D reconstructions of SRXTM images for: \textit{(a)} CAST2-1; \textit{(b)} CAST2-6; \textit{(c)} CAST2-7; \textit{(d)} CAST2-12 samples. Inclusions are clearly visible in brighter colors since composed by chemical elements heavier than carbon. Estimates of inclusions and diamonds sizes, expressed in \si{\um} units, are shown for arbitrary fixed sample orientations in orange and light-blue color, respectively.}
\label{fig:tomo-images}
\end{figure}

3D reconstructions of the tomographic projections for the four diamonds selected for magnetic characterization are shown in \figurename~\ref{fig:tomo-images}. Estimates of the linear sizes of both inclusions and diamonds are reported for some arbitrarily chosen sample orientation. From these estimates we can conclude that the inclusion size ranges approximately between \SI{100}{\um} and \SI{700}{\um}, while the diamond size is $\sim\,$\SIrange[range-phrase=--]{0.5}{1.5}{\mm}, in agreement with the outcomes of optical microscopy analysis (see \figurename~\ref{fig:CAST2-diamonds}). A careful analysis of 2D tomographic slices through the reconstructed volume of the four diamonds allowed us to conclude that only the CAST2-1 inclusion is fully entrapped into the diamond, while the inclusions in the other three diamonds are connected to the external surroundings being therefore epigenetic with respect to the host. Another possible explanation for the appearance of fractures can arise whenever the inclusions have higher compressibility with respect to the host and thus expand in volume more than diamond, when travelling towards the Earth's surface. In such a case, inclusions may also be proto- or syngenetic with respect to the host diamond even in presence of fractures.

\subsection{\label{subsec:AGFM-res} AGFM}
\begin{figure}[htbp]
\centering
\includegraphics[width=0.47\textwidth]{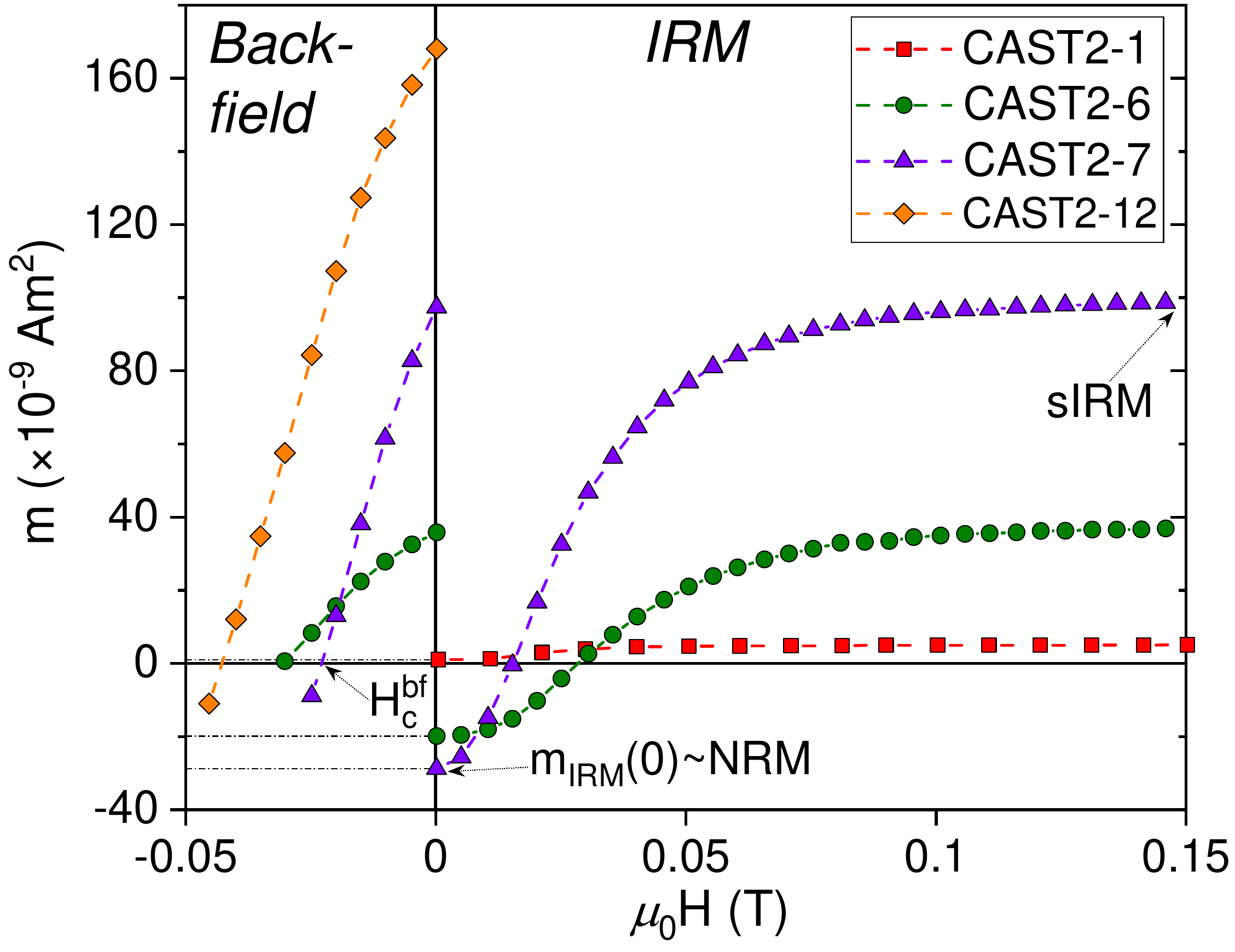}
\caption{IRM ($H\geq 0$) and backfield ($H\leq 0$) curves, acquired with the AGFM, on CAST2-1 (red squares), CAST2-6 (green circles), CAST2-7 (violet triangles) and CAST2-12 (orange diamonds) samples. Points describing the remanence at zero field $m_\text{IRM}(0)$ (best approximation of the NRM), the remanence at saturation sIRM, reached at \SI{0.15}{\tesla} for all samples, and the coercive field of remanence $H_\text{c}^\text{bf}$ are indicated with arrows; corresponding values are reported in \tablename~\ref{tab:IRM-bf-hyst-values}.}
\label{fig:IRM-bf-exp}
\end{figure}

IRM curves have been first acquired on CAST2-1, -6, -7 samples, while backfield curves have been later collected on CAST2-6, -7 and -12 samples. Resulting curves are shown in \figurename~\ref{fig:IRM-bf-exp}, while $m_\text{IRM}(0)$ and saturation $m_\text{IRM}$ (sIRM) values, the latter reached for all samples at \SI{0.15}{\tesla}, are listed in \tablename~\ref{tab:IRM-bf-hyst-values}. The same table reports also the values of the coercivity of remanence $H_\text{c}^{\text{bf}}$, which identify on each backfield curve the magnetic fields for which $m_\text{bf}(H_\text{c}^\text{bf})=0$.

\begin{table*}
\centering
\begin{tabular}{l*{2}{>{\centering}p{2.35cm}}>{\centering}p{1.2cm}*{2}{>{\centering}p{2.35cm}}>{\centering\arraybackslash}p{1.2cm}}
\toprule
\multirow{3.1}{*}{\textbf{Sample}} & \multicolumn{3}{c}{\textbf{IRM/backfield parameters}} & \multicolumn{3}{c}{\textbf{Hysteresis parameters}} \\
\cmidrule(lr){2-4} \cmidrule(lr){5-7}
 & $\boldsymbol{m_\text{\textbf{IRM}}(0)}$ & \textbf{sIRM} & $\boldsymbol{\mu_0 H_\text{\textbf{c}}^\text{\textbf{bf}}}$ & $\boldsymbol{m_\text{\textbf{sat}}}$ & $\boldsymbol{m_\text{\textbf{r}}}$ & $\boldsymbol{\mu_0 H_\text{\textbf{c}}}$ \\
& \textbf{[$\boldsymbol{\times}$\SI[detect-all=true]{e-9}{\ampere\metre\squared}]} & \textbf{[$\boldsymbol{\times}$\SI[detect-all=true]{e-9}{\ampere\metre\squared}]} & \textbf{[\si[detect-all=true]{\tesla}]} & \textbf{[$\boldsymbol{\times}$\SI[detect-all=true]{e-9}{\ampere\metre\squared}]} & \textbf{[$\boldsymbol{\times}$\SI[detect-all=true]{e-9}{\ampere\metre\squared}]} & \textbf{[\si[detect-all=true]{\tesla}]} \\
\midrule
CAST2-1 & 1 & 5 & // & 27 & 5 & 0.006 \\
\cmidrule(lr){2-4} \cmidrule(lr){5-7}
CAST2-6 & -20 & 37 & -0.031 & 112 & 35 & 0.020 \\
\cmidrule(lr){2-4} \cmidrule(lr){5-7}
CAST2-7 & -29 & 99 & -0.023 & 541 & 98 & 0.012 \\
\cmidrule(lr){2-4} \cmidrule(lr){5-7}
CAST2-12 & 60 & 168 & -0.043 & 510 & 174 & 0.026 \\
\bottomrule
\end{tabular}
\caption{Numerical values of: IRM remanence at zero field ($m_\text{IRM}(0)$), IRM at saturation field \SI{0.15}{\tesla} (sIRM), IRM coercive field ($H_\text{c}^\text{bf}$), as extrapolated from IRM/backfield curves (\figurename~\ref{fig:IRM-bf-exp}); saturation magnetic moment ($m_\text{sat}$), remanent magnetic moment ($m_\text{r}$) and coercive field ($H_\text{c}$), as extrapolated from hysteresis loops (\figurename~\ref{fig:hyst-exp}), for the CAST2 diamonds reported in the first column.}
\label{tab:IRM-bf-hyst-values}
\end{table*}

\begin{figure}[htbp]
\centering
\includegraphics[width=0.47\textwidth]{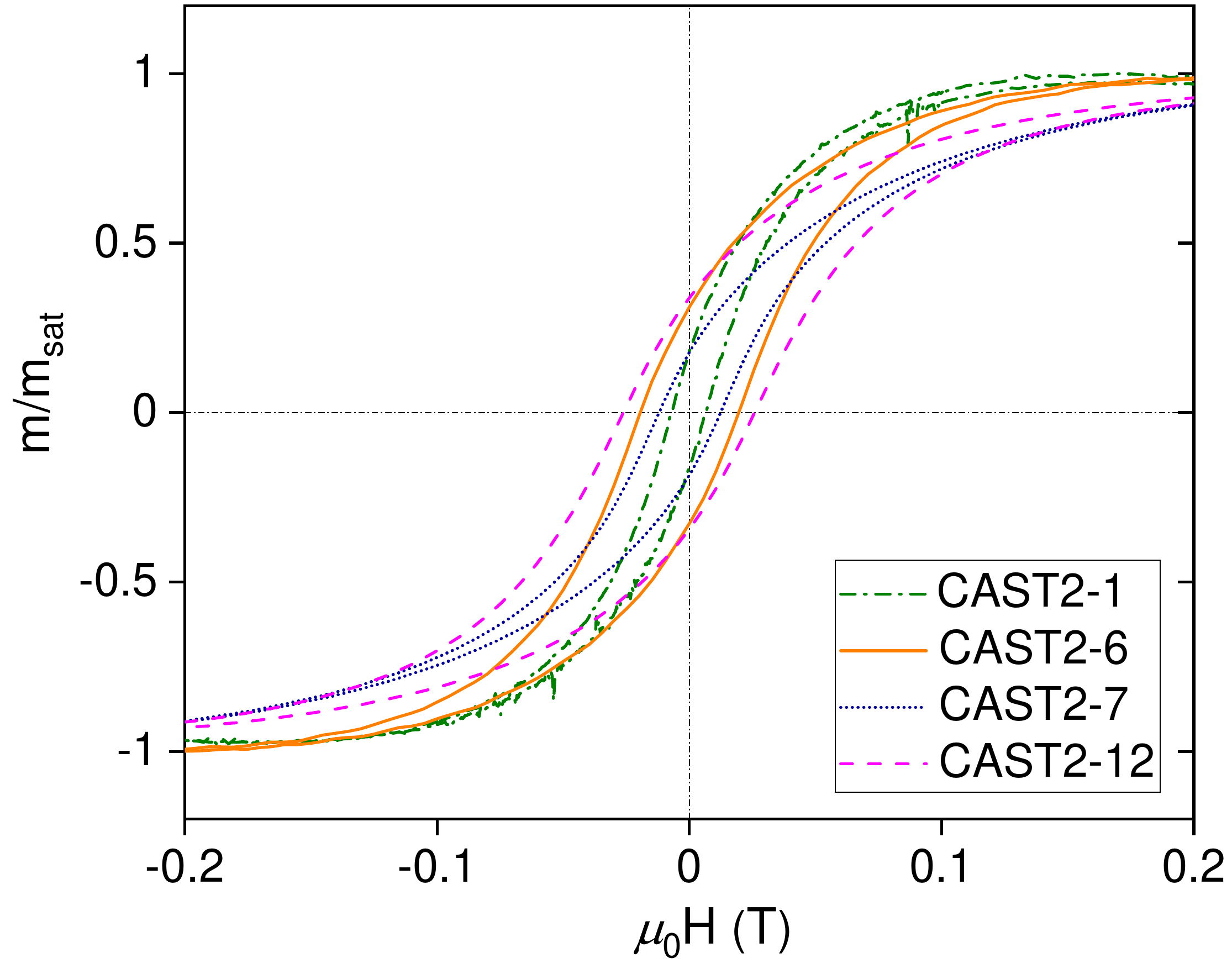}
\caption{Hysteresis loops, acquired with the AGFM, on CAST2-1 (dash-dotted, green), CAST2-6 (solid, orange), CAST2-7 (dotted, blue) and CAST2-12 (dashed, pink) samples. The dimensionless magnetic moment is obtained by diving $m$ by $m_\text{sat}$ values reported in \tablename~\ref{tab:IRM-bf-hyst-values}. The diamagnetic contribution of diamonds, becoming relevant at $\mu_0 H\gg$ \SI{0.2}{\tesla} (CAST2-1, -6 samples) or \SI{0.5}{\tesla} (CAST2-7, -12 samples), is not shown. A sample-dependent vertical offset due to instrumental noise has been subtracted from the curves.}
\label{fig:hyst-exp}
\end{figure}

After IRM and backfield curves, major hysteresis loops reporting the detected magnetic moment $m$ as a function of the uniform applied field $H$ have been acquired for all the four samples and are displayed in \figurename~\ref{fig:hyst-exp}. In the loops, the magnetic moment is expressed in dimensionless units by dividing the scalar component $m$ measured with the AGFM by the saturation moment $m_\text{sat}$, which represents the value of $m$ reached at the saturation field $H_\text{sat}$ introduced in Sec.~\ref{subsec:AGFM}. For $H\geq H_\text{sat}$, hysteresis loops show a characteristic plateau and no further magnetization processes due to domain wall motion, spin rotation or spin reversal, occur anymore so that $m$ remains essentially constant. In the present case, $m_\text{sat}$ values have been chosen as the maximum reached by $m$ for each sample and are reported in \tablename~\ref{tab:IRM-bf-hyst-values}. The values of the remanent magnetic moment $m_\text{r}=m(H=0)$ and of the coercive field $H_\text{c}$ at which $m(H_\text{c})=0$, as extrapolated from the curves, are also reported in \tablename~\ref{tab:IRM-bf-hyst-values}. We notice that sIRM and $m_\text{r}$ values can be considered equal within the uncertainty value, as expected since sIRM can be identified with the usual remanence associated to a major complete hysteresis loop of a magnetic material.

\begin{figure}[htbp]
\centering
\subfloat[][]{\includegraphics[width=0.247\textwidth]{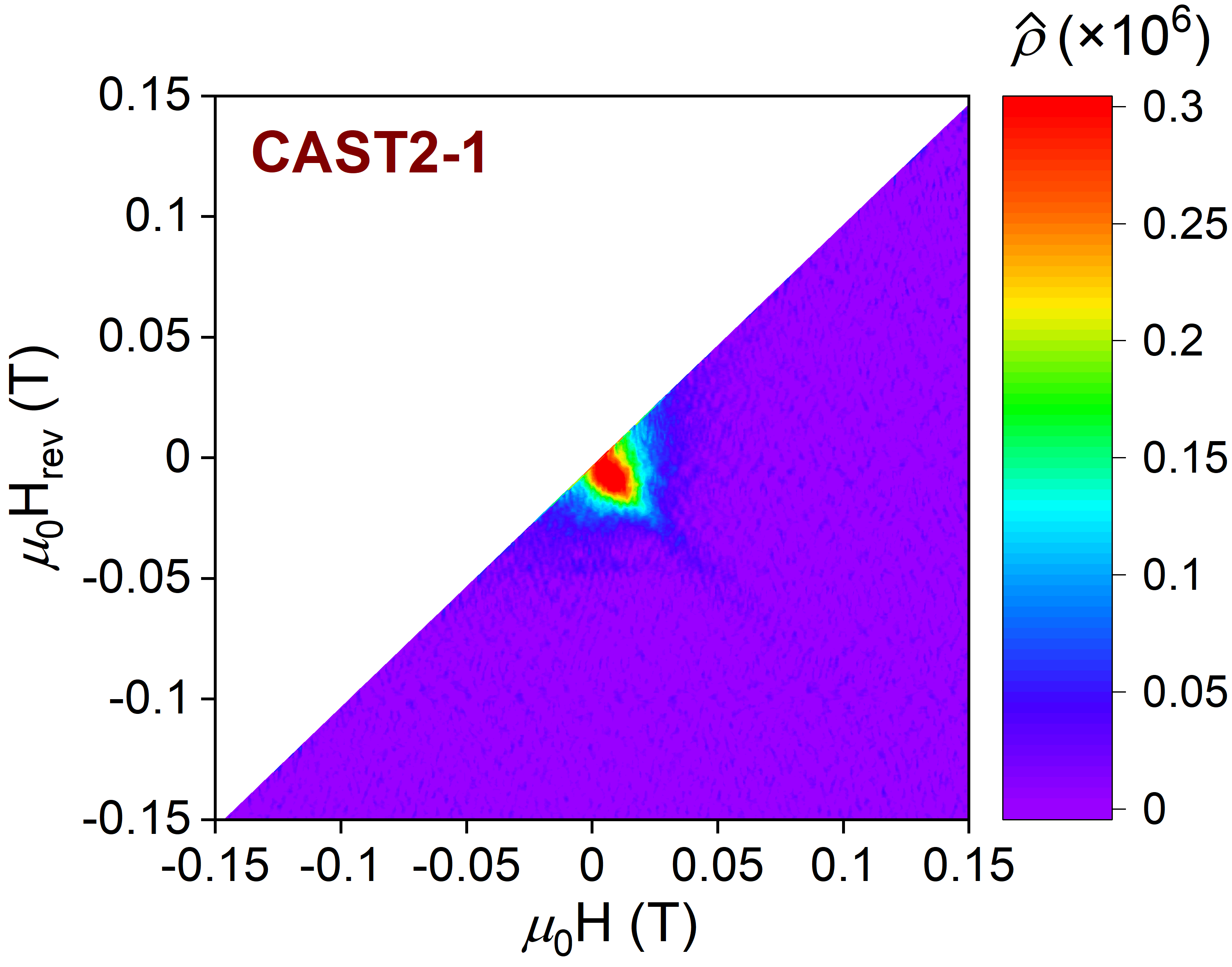}}
\subfloat[][]{\includegraphics[width=0.244\textwidth]{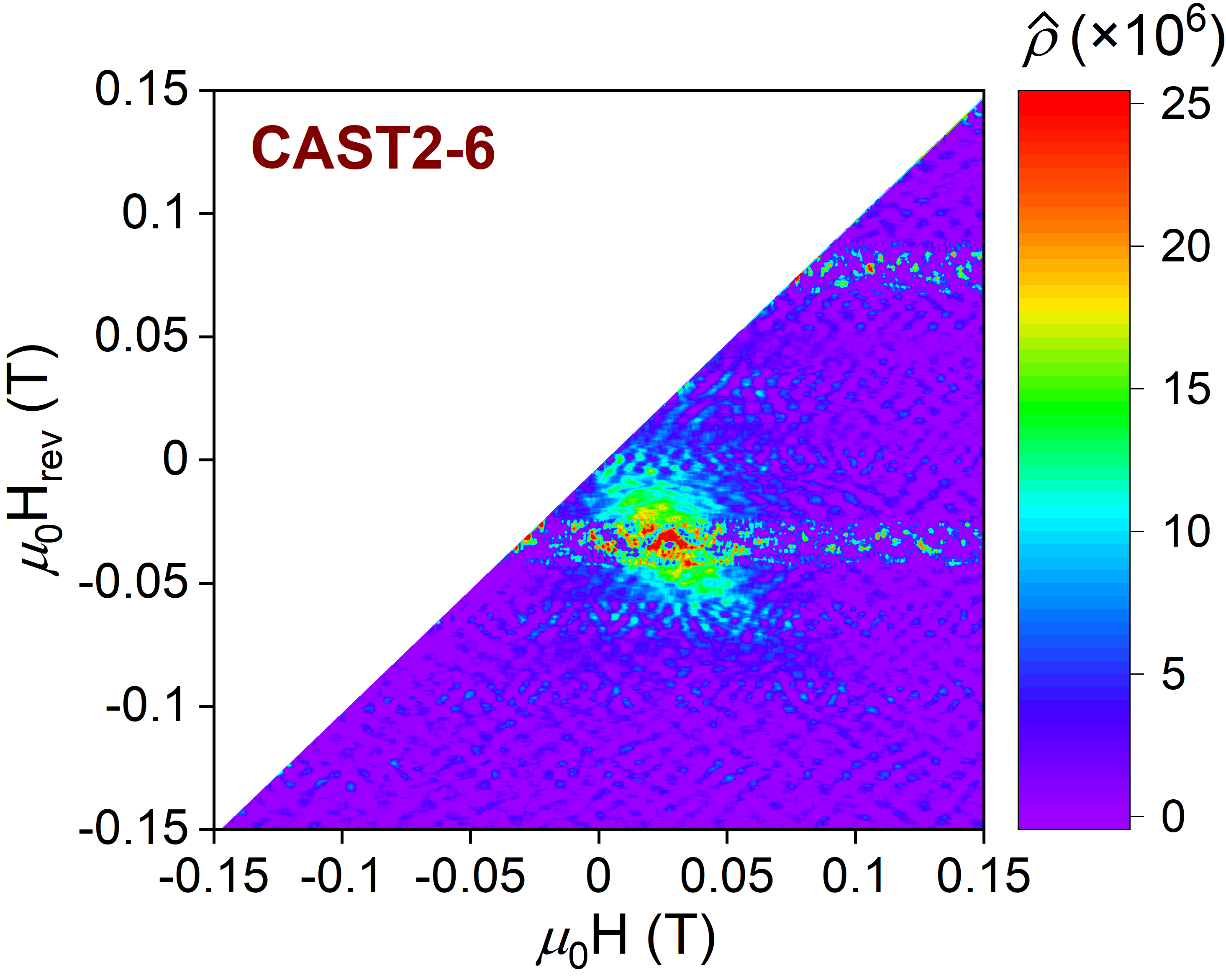}}\\
\subfloat[][]{\includegraphics[width=0.246\textwidth]{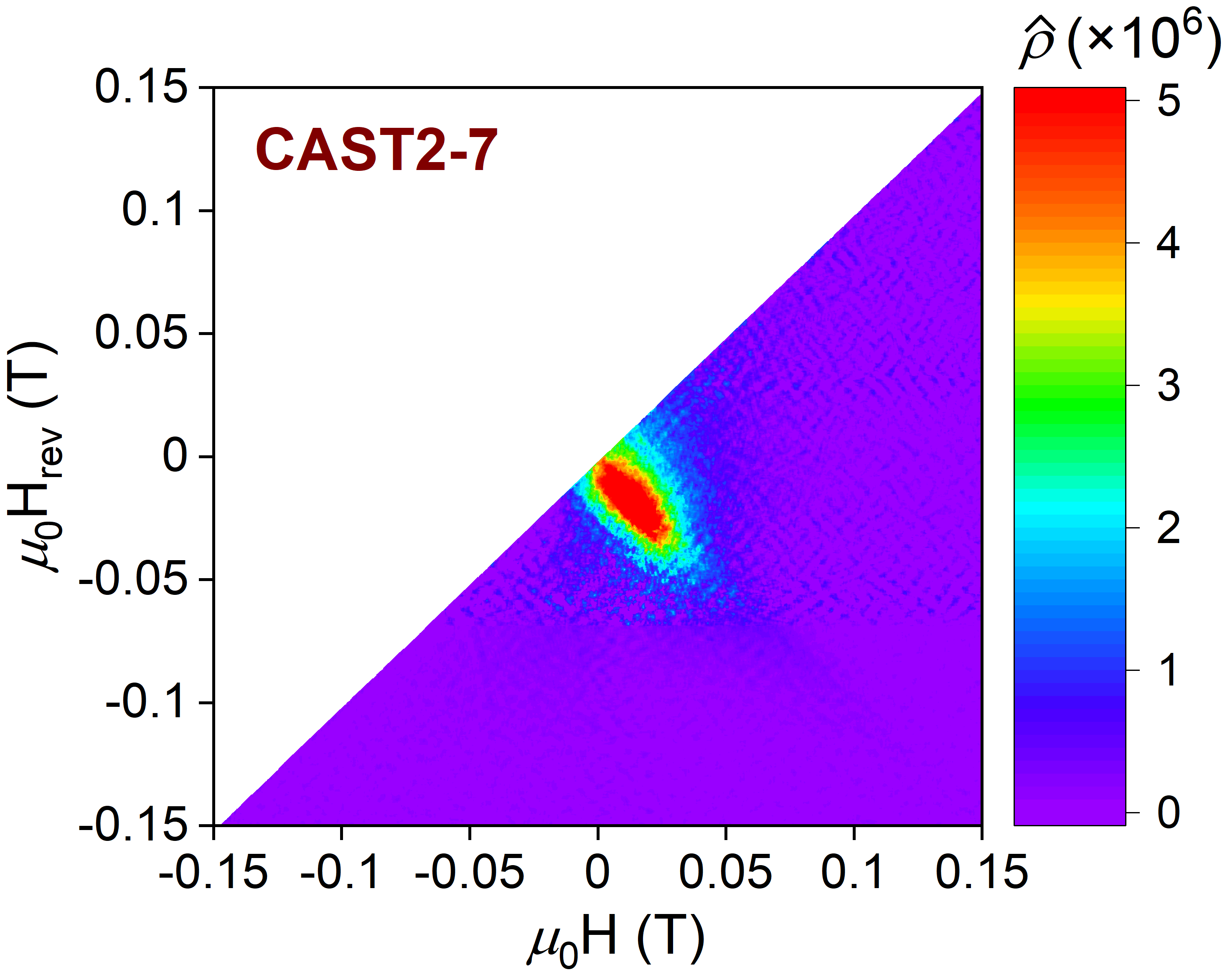}}
\subfloat[][]{\includegraphics[width=0.246\textwidth]{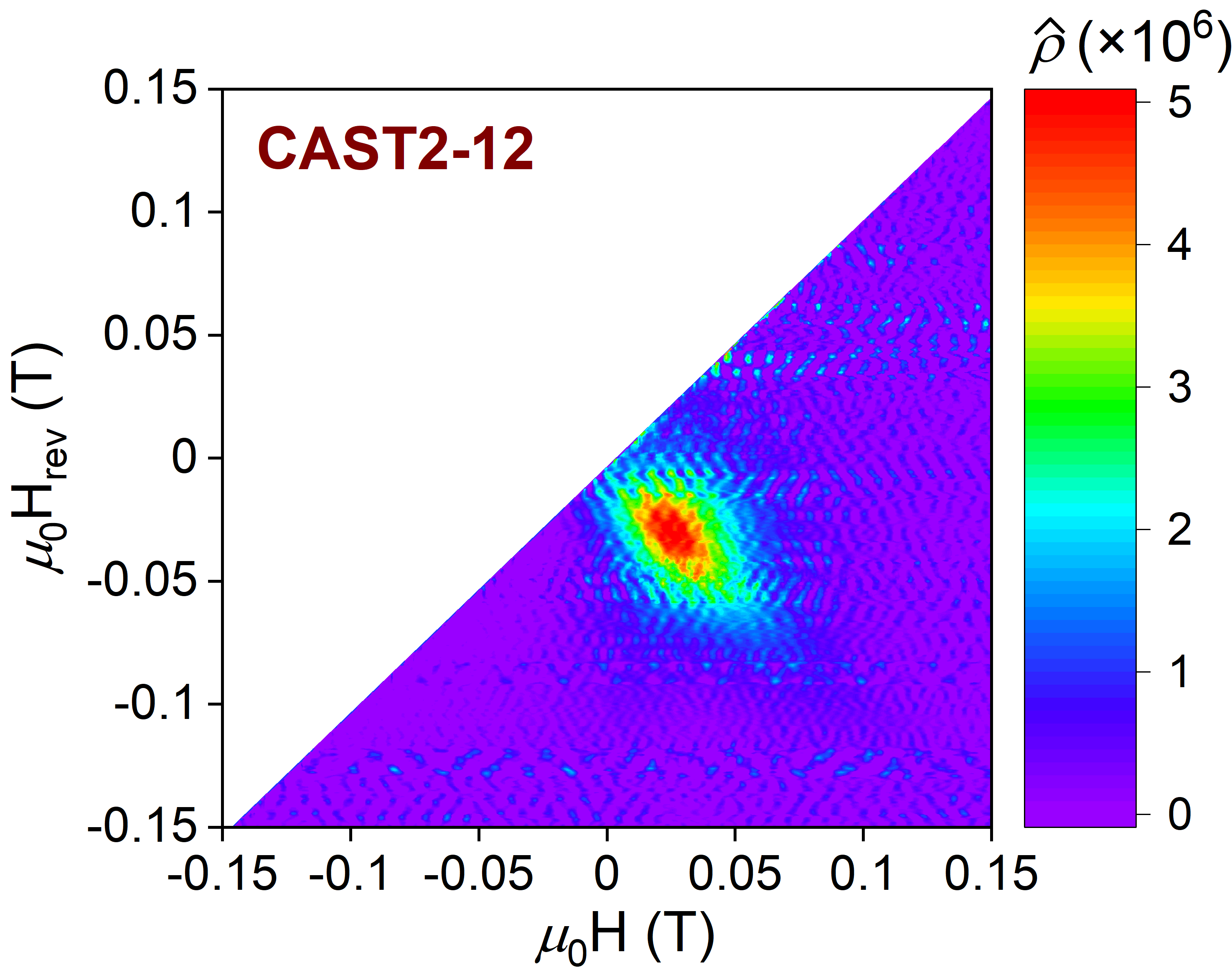}}
\caption{2D FORC diagrams evaluated from full sets of FORCs acquired with the AGFM on: (\textit{a}) CAST2-1; (\textit{b}) CAST2-6; (\textit{c}) CAST2-7; (\textit{d}) CAST2-12 samples. FORC distribution values are expressed in dimensionless units as $\hat{\rho}=(H_0^2/m_0)\rho$, where $\rho$, $H_0$ and $m_0$ are defined in Sec.~\ref{subsec:AGFM-res}. Reversible contribution to $\hat{\rho}$, lying on the $H-H_\text{rev}=0$ bisector, is not included.}
\label{fig:FORC-diagrams}
\end{figure}

Finally, full sets of FORCs have been collected for the four samples as detailed in Sec.~\ref{subsec:AGFM} and after their acquisition FORC diagrams have been evaluated as 2D contour plot of the 3D function $\rho(H,H_\text{rev})=-\partial^2 m/\left(\partial H\partial H_\text{rev}\right)$, known as FORCs distribution. For the evaluation of $\rho$ various numerical methods have been developed, as described in Ref.~\cite{Pike-FORCs,Roberts-FORCs}. Resulting 2D FORC diagrams are reported in \figurename~\ref{fig:FORC-diagrams} where, for a better comparison, $\rho$ values have been expressed in dimensionless units as $\hat{\rho}(H,H_\text{rev})=(H_0^2/m_0)\rho(H,H_\text{rev})$. In the previous relation $H_0$ is a characteristic magnetic field that, for all samples, we set equal to the saturation field of the IRM curves, i.e. $\mu_0 H_0=$ \SI{0.15}{\tesla}, while $m_0$ is a characteristic magnetic moment that we fixed equal to the saturation value $m_\text{sat}$ of the hysteresis loop of each sample reported in \tablename~\ref{tab:IRM-bf-hyst-values}. Finally, we notice that the reversible contribution to $\hat{\rho}$ due to the points $m(H_\text{rev};H_\text{rev})$, lying on the descending branch of the major hysteresis loop, is not included into the diagrams. Indeed, $\hat{\rho}$ is correctly evaluated only for $H>H_\text{rev}$ because its definition involves second derivatives, and hence the reversible contribution to $\hat{\rho}$ must be added, when necessary, by making proper ansatz about its analytical behaviour \cite{Pike-FORCs-rev}.  

\section{\label{sec:discussion} Discussion}
The combination of XRD, SRXTM and magnetic data we acquired allows to develop the following crystallographic, inner-structural  and physical picture about the inclusions present in the four diamonds we investigated. All samples contain one magnetic phase, in some cases comprised in more than one inclusion for each diamond, that XRD allows to determine as a polycrystalline iron oxide, restricting the possibilities to magnetite or magnesioferrite. Furthermore, microtomography shows that only the CAST2-1 diamond has a fully entrapped and isolated inclusion, while the other samples have fractures connecting the external surface to the embedded magnetic oxides. Therefore, CAST2-1 is the only sample that may comprise a proto- or syngenetic inclusion, although its sheet-like shape is similar to that of the inclusions found in the fractured diamonds (\figurename~\ref{fig:tomo-images}). The knowledge of the saturation magnetic moment of the samples $m_\text{sat}$ from AGFM measurements, combined with estimates of the volume of the magnetic inclusions $V_\text{mag}$ as extrapolated from SRXTM data, can then provide useful insights to distinguish between magnetite and magnesioferrite phases. Indeed, these two Fe-rich spinels have different saturation magnetization $M_\text{sat}=m_\text{sat}/V_\text{mag}$ due to the substitution of Mg for Fe$^{2+}$ in the octahedral B sites and, partially, for Fe$^{3+}$ in the tetrahedral A sites of magnetite \cite[pp.~178--180]{Cullity-book}. For the CAST2-1 sample, the analysis of tomographic data has shown that the volume of the diamond alone is $V_\text{diam}\simeq$ \SI{1227.8e-13}{\metre\cubed}, the volume of non-magnetic phases possibly present in negligible amount inside the diamond and of fractures is $V_\text{fract}\simeq$ \SI{3.2e-13}{\metre\cubed}, while the magnetic phase comprised in the inclusion has volume $V_\text{mag}\simeq$ \SI{0.6e-13}{\metre\cubed}. This means that the volume of the whole sample is $V_\text{CAST2-1}=V_\text{diam}+V_\text{fract}+V_\text{mag}\simeq$ \SI{1231.6e-13}{\metre\cubed} and that the magnetic inclusion volume $V_\text{mag}$ is about $0.05\%$ of $V_\text{CAST2-1}$. By combining $V_\text{mag}$ with the $m_\text{sat}=$ \SI{27e-9}{\ampere\metre\squared} value reported in \tablename~\ref{tab:IRM-bf-hyst-values}, we obtain an estimate of the saturation magnetization at room temperature for the CAST2-1 inclusion which is $M_\text{sat}\simeq$ \SI[per-mode=symbol]{486e3}{\ampere\per\metre}. Since the reported $M_\text{sat}$ values at \SI{293}{\kelvin} for magnetite and magnesioferrite are \SI[per-mode=symbol]{480e3}{\ampere\per\metre} and \SI[per-mode=symbol]{120e3}{\ampere\per\metre} respectively \cite[p.~183]{Cullity-book}, we can conclude that CAST2-1 inclusion is most probably magnetite. Similar reasoning may be applied to the inclusions within the other samples in order to identify more precisely the magnetic phase composing them.

Magnetic signals detected with the AGFM by collecting IRM/backfield curves, hysteresis loops and FORCs show a quite complex behaviour making the development of a fully comprehensive physical interpretation quite difficult. IRM curves (\figurename~\ref{fig:IRM-bf-exp}) remarkably show that $m_\text{IRM}(0)$ values are different from zero in all the samples, meaning that they all carry a detectable NRM. This important outcome, holding as long as the exposure to magnetic fields other than the Earth's one $H_\text{GMF}$ can be ruled out as in our case, can provide interesting information on the amplitude of $H_\text{GMF}$ at the time of formation of the inclusions, since the time history of $H_\text{GMF}$ influences the NRM recorded by the samples. It is worth pointing out that information about the declination and inclination of $H_\text{GMF}$ is instead not available from the kind of inclusions and measurements here proposed, because of the random orientation that the diamonds we have investigated had in the alluvial valley at the time of their eruption and subsequent extraction. As a second remark, $m_\text{IRM}(0)$ and sIRM values span more than one order of magnitude. Indeed, $m_\text{IRM}(0)$ is about \SI{e-9}{\ampere\metre\squared} for CAST2-1 sample and of the order of \SI{e-8}{\ampere\metre\squared} for CAST2-6 and CAST2-7 samples. The same circumstance occurs for sIRM values, that vary between $\sim\,$\SI{5e-9}{\ampere\metre\squared} for CAST2-1 and $\sim\,$\SI{1e-7}{\ampere\metre\squared} for CAST2-7 samples. Various reasons may explain the observed quite large variations. The most obvious one is that inclusions comprise different magnetic materials, but it does not apply to the samples here investigated because of the XRD results. Another possibility is that magnetic inclusions vary in size and volume from sample to sample and microtomographic images (\figurename~\ref{fig:tomo-images}) show that we can rely on these differences to partially explain the different outcomes in the IRM behaviour. A final possibility, applying as well in our case, can be that the magnetic moment of a specimen is a vector quantity, but the AGFM is able to perform only scalar measurements. Understanding which of the last two reasons play a major role in each sample we have investigated is a challenging task deserving specific consideration in future works.

Hysteresis loop reflect the main properties of a magnetic system, as its magnetic anisotropy, magnetic susceptibility, coercive forces and saturation magnetization \cite{Cullity-book, Bertotti-hyst}, while the shape of FORCs diagrams shed light on the microscopic magnetic configuration of the particles composing a system, which is particularly interesting when dealing with natural samples as mineral inclusions in host diamonds or rocks \cite{Pike-FORCs,Roberts-FORCs}. In particular, hysteresis is a complex phenomenon strictly related in our samples to the their magnetic granulometry, i.e. the relative distribution of superparamagnetic, single-domain, or multi-domain magnetic particles according to the composition, size and geometry of the included Fe-based crystallites (magnetite or magnesioferrite), whose associated magnetization vectors can have varied amplitudes pointing in different directions with relaxation times mainly depending on their size. Detailed descriptions of such processes can be found in many textbooks devoted to the subject \cite{Bertotti-hyst,Dunlop-book}. The loops we acquired on our samples (\figurename~\ref{fig:hyst-exp}) can offer a qualitative information about the magnetic anisotropy and the microgranulometry of the inclusions, encompassed in their behaviour close to the saturation field $H_\text{sat}$ and to the coercive field $H_\text{c}$. The behaviour of the loops close to saturation is known to be closely dependent on an intrinsic factor, the magnetic anisotropy, with systems having lower uniaxial anisotropy constant $K$ being characterized by lower $H_\text{sat}$ values (see \cite[pp.~218--222]{Cullity-book}~\cite{StonerWohlfarth-model} for a detailed explanation). According to this general observation and by assuming to deal with inclusions characterized by an effective uniaxial anisotropy constant $K_\text{eff}$, encompassing both the effects of their crystal structure and of the strains, dislocations, defects induced by the host diamond, we can conclude that CAST2-1 and CAST2-6 diamonds ($\mu_0 H_\text{sat}\simeq$ \SI{0.2}{\tesla}) shall most probably have lower $K_\text{eff}$ with respect to CAST2-7 and CAST2-12 samples ($\mu_0 H_\text{sat}\simeq$ \SI{0.5}{\tesla}). Similarly, the behaviour of the hysteresis close to $H_\text{c}$ gives insights about an extrinsic factor which is the grain size of the magnetic particles composing the inclusions. In this case, it is known that the higher the coercivity is, the lower the grain size of the particles and more crystal defects are most probably present in the system \cite[pp.~360--364]{Cullity-book}~\cite{Luborsky-coercivity,Kneller-coercivity}. Then, by looking at the $H_\text{c}$ values reported in \tablename~\ref{tab:IRM-bf-hyst-values}, we can conclude that CAST2-6 and CAST2-12 inclusions shall comprise finer grains with respect to CAST2-1 and CAST2-7 samples, respectively. To get more quantitative results, models of hysteresis must be developed to relate $H_\text{c}$ values to extrinsic parameters such as the grain size of the particles.   

The symmetry of the FORC diagrams (\figurename~\ref{fig:FORC-diagrams}) around the $H_\text{rev}+H=0$ axis is to be expected for magnetic systems, due to the symmetry of ascending and descending branches of the hysteresis loops around the origin $(H=0,M=0)$, i.e. $M_\text{desc}(-H)=-M_\text{asc}(H)$ with $M_\text{asc}$ ($M_\text{desc}$) being the magnetization value evaluated on the ascending (descending) branch of the loop. The shape of the diagrams, appearing in all the cases slightly spread out in the $H-H_\text{rev}=0$ direction, is due to the presence of single-domain magnetic particles with not negligible local interactions among them, as exhaustively explained in Ref.~\cite{Pike-FORCs,Roberts-FORCs}. It is worth mentioning that when the particles crystallize at different times, the presence of such coupling makes it more difficult to establish at which point the NRM they carry can be ascribed to the action of the Earth's magnetic field. In particular, we see from \figurename~\ref{fig:FORC-diagrams} that local coupling plays a bigger role in CAST2-6 and CAST2-12 inclusions with respect to CAST2-1 and CAST2-7 ones, since the contour plots of the former samples show less sharp peaks. It is noteworthy that FORC diagrams did not evidence negative peaks usually associated to interactions among different magnetic objects, or inclusions, in the analyzed systems. Finally, the position of the $\rho$ distribution peaks, defined as the point $(\bar{H},\bar{H}_\text{rev})$ at which the contour plot reaches its maximum, provides an estimate of the mean value $\bar{H}_\text{c}$ of the distribution of coercive fields and energy barriers that are associated to each magnetic particle of the inclusion. By expressing the peak position as $H_\text{peak}=\left(\bar{H}-\bar{H}_\text{rev}\right)/2$, we can again gather the samples into two groups: on the one side we have CAST2-1 and CAST2-7 samples, having $\mu_0 H_\text{peak}\simeq$ \SI{0.006}{\tesla} and \SI{0.017}{\tesla} respectively, while on the other side there are CAST2-6 and CAST2-12 samples, with peaks values at $\mu_0 H_\text{peak}\simeq$ \SI{0.029}{\tesla} and \SI{0.026}{\tesla} respectively. The difference in $H_\text{peak}$ values can be ascribed in our case to variations in the grain size and in the orientation of the magnetization within the particles belonging to the various samples. The latter explanation and the extrapolated $\bar{H}_\text{c} \sim H_\text{peak}$ values are is in agreement with the $H_\text{c}$ estimates reported in \tablename~\ref{tab:IRM-bf-hyst-values}.

\section{\label{sec:conclusions} Conclusions}
In the present paper we have proposed an efficient, non-destructive experimental methodology to determine the inner structure, the crystallographic and the magnetic properties of inclusions entrapped in a series of natural diamonds. The methodology is based on the use of XRD, SRXTM and AGFM techniques and has the potential to be successfully applied to any host-inclusion system of \si{\mm}/sub-\si{mm} size comprising magnetic phases with magnetic moments higher than $\sim\,$\SI{e-9}{\ampere\metre\squared}. It allows to build up a qualitatively comprehensive and robust picture of the main chemico- and geophysical features of the samples under investigation.

In particular, for the four diamonds we selected in our study, we have shown that magnetic inclusions comprise polycrystalline iron oxides having ferrimagnetic spinel structure identified as magnetite or magnesioferrite. We have demonstrated the presence of fractures connecting the inclusions to the diamond surface in all but one samples, thus suggesting that this kind of inclusions are epigenetic and probably formed because of pre-existing Fe-rich fluids that percolated through the cracks and diffused within diamonds. We have shown that all the inclusions carry a non zero NRM and we have interpreted the rich picture arising from the FORC diagrams by assuming that the inclusions are composed of locally interacting single-domain particles. Finally, we have distinguished the samples according to the different behaviour observed in their hysteresis loops, by relating the latter to both intrinsic and extrinsic factors represented by the magnetic anisotropy and the grain size of the particles. The comparison of experimental data with proper models of magnetic hysteresis is needed to derive more quantitative conclusions on this topic. 
 
\begin{acknowledgments}
This project has received funding from the European Research Council (ERC) under the European Union's Horizon 2020 research and innovation programme (grant agreement No. 714936 for the project TRUE DEPTHS to M. Alvaro). MA has also been funded by the FARE MIUR n. R164WEJAHH for the project IMPACt. 

We acknowledge the Paul Scherrer Institut, Villigen, Switzerland for provision of synchrotron radiation beamtime at the TOMCAT beamline of the SLS.
\end{acknowledgments}

\bibliography{Piazzi-et-al-biblio-Diam-rel-mater-2019}

\end{document}